\documentclass[english]{article}
\usepackage{graphicx,color}
\def\no{\noindent}
\def\bc{\begin{center}}
\def\ec{\end{center}}
\def\vs{\vskip0.5cm}
\def\beq{\begin{equation}}
\def\eeq{\end{equation}}

\def\br{{\bf r}}

\def\bq{{\bf q}}

\def\bQ{{\bf Q}}

\def\be{{\bf e}}
\def\bv{{\bf v}}

\begin{document}

\title{
Localized states in monitored quantum walks 
}

\author{K. Ziegler\\
Institut f\"ur Physik, Universit\"at Augsburg\\
D-86135 Augsburg, Germany
}

\maketitle

\no
Abstract:

In this paper we study localized states in a monitored evolution
on a finite graph and how they are distinguished from the delocalized states 
in terms of the transition probabilities and the mean transition times.
Monitoring is performed by repeated projective measurements with respect
to a single quantum state.
Our constructive approach is based on a mapping from a set of energy levels and
an eigenvector basis onto the monitored evolution matrix. The eigenvalues
of the latter are distributed over the complex unit disk and the corresponding
transition probabilities decay quickly in the quantum Zeno regime at
frequent measurements.
A localized basis favors the return to the initial state, while a delocalized
basis favors transitions between different states.
This provides a practical criterion to identify localized 
states by measuring the mean transition time.



\section{Introduction}


Transport properties of quantum systems distinguish between localized
and delocalized regimes. For instance, in the delocalized regime a particle may travel over
large distances and explore the entire Hilbert space, while in the localized regime it stays for
all times close to the position where it was initially created. This behavior is 
determined by the eigenstates of the evolution operator and is reflected
by the transition probability, where the latter is a classical quantity. The example of the particle position
can be generalized to other quantum states, always asking whether or not
an initial state can explore the entire Hilbert space under quantum
evolution or only a subspace. A spin-ordered state, for example, keeps
its order in the localized regime, while it may loose it over time
in the delocalized regime.

The appearance of localized states and the competition with its delocalized counterparts 
is a crucial effect in disordered systems.
Disorder usually means that the ensemble of eigenstates is a mixture of localized as well as
delocalized states. The distribution of the two types of states has different weights, in which one
of them wins and dominates the average result. 
This has been studied intensely for the unitary 
evolution~\cite{ande.58,ab.an.79,doi:10.1142/7663,stolz2011introduction}, indicating that it is worthwhile 
to study these properties also in terms of monitored quantum walks. 
%
In any case, the disorder average is performed on a classical quantity, e.g., the transition
probability, while quantum average is always taken before. 
Therefore, for any realization of a random Hamiltonian we first determine its eigenbasis,
calculate the quantum expectation value of the transition probability and average
with respect to the distribution of the Hamiltonians only in the final step.
In the following we will focus on the first two steps.

Quantum walks on finite 
graphs~\cite{Aharonov1993,2003ConPh..44..307K,PhysRevLett.102.180501,MULKEN201137,Das_2022} 
offer a systematic approach to a deeper analysis of a quantum evolution by constructing specific
evolution operators and study the corresponding transition properties. In particular, we 
can employ repeated projective measurements to monitor the localization properties.
A monitored evolution, in contrast to a unitary evolution, provides an efficient approach to determine
properties of the quantum system. For instance, the mean transition time reveals after how many
measurements the systems typically has performed a transition from the initial state to another state 
of the quantum system.
Very frequent measurements result in the
Quantum Zeno Effect~\cite{Misra,KOSHINO2005191,facchi08,2020arXiv200801070B}.
It competes with the localization effect, since both cause the quantum walk
to stay near the initial state. 

The goal of this paper is to study the role of localized versus delocalized
eigenstates in a monitored quantum evolution. To this end, we briefly discuss
the unitary evolution and evaluate the time-averaged transition probability on
a finite graph and its properties for localized as well as for delocalized states.
%
Central to our systematic approach is that we map the continuous-time 
unitary evolution of a quantum system
onto a discrete-time evolution through a specific measurement protocol. This concept is quite
useful for practical realizations on a quantum computer~\cite{PhysRevResearch.5.033089}. 
It has some similarity with the coin flip induced discrete-time 
quantum walks~\cite{Aharonov1993,2003ConPh..44..307K}. 
But instead of a coin flip we employ here a projective measurement to create a 
discrete quantum walk.
%
We describe the evolution of a quantum system in terms of a measurement protocol,
which requires typically only a finite number of measurements for a complete description
on a finite-dimensional Hilbert space.
%
In this context
spectral degeneracies are special cases, which may require in principle an infinite
number of measurement for this protocol, since the transition probabilities 
do not decay~\cite{krovi06,krovi06a}.
On the other hand, they are crucial for phase transitions, where the energy levels of two
or more different groundstates cross. 

We construct an ensemble of Hamiltonians with localized and with delocalized eigenvectors.
In the simplest case we consider just a Hamiltonian with localized and another Hamiltonian with 
delocalized eigenvectors. 
%
Measurements are defined as local operations. Since localized states induce
a local structure, it is interesting to see how these two local effects compete and interact. 
Our approach is based on the mapping from a given spectrum (i.e., given energy levels) and a specific
set of eigenstates onto an evolution operator. 
This mapping reflects the situation of the experiment, where we 
typically observe energy levels rather than Hamiltonians or evolution operators. 
Recent studies are based on quantum circuits, which consist of unitary operators without 
referring to a Hamiltonian. In those cases the construction of the quantum circuit relies
on the eigenvalues and the eigenvectors of the evolution operators.  

The paper is organized as follows. 
In Sect. \ref{sect:properties} the monitored evolution, based on repeated projective measurements,
is discussed and compared with the unitary evolution. Localized states are defined through the inverse
participation ratio and the monitored transition amplitude is given in terms of quantum walk sums
in Sect.~\ref{sect:loc_states}.
Finally, a recursive approach for the transition amplitude is briefly described in Sect. \ref{sect:recursion}. 
The results of the general part are discussed in Sect. \ref{sect:discussion} and the
inverse participation ratio, the transition amplitude and the mean transition times are calculated for 
several examples.
The derivation of some relations and the details of the calculations are collected in 
App. \ref{app:qws} -- \ref{app:local_basis}.


\begin{figure}[t]
\centering
\includegraphics[width=3.5cm,height=5cm]{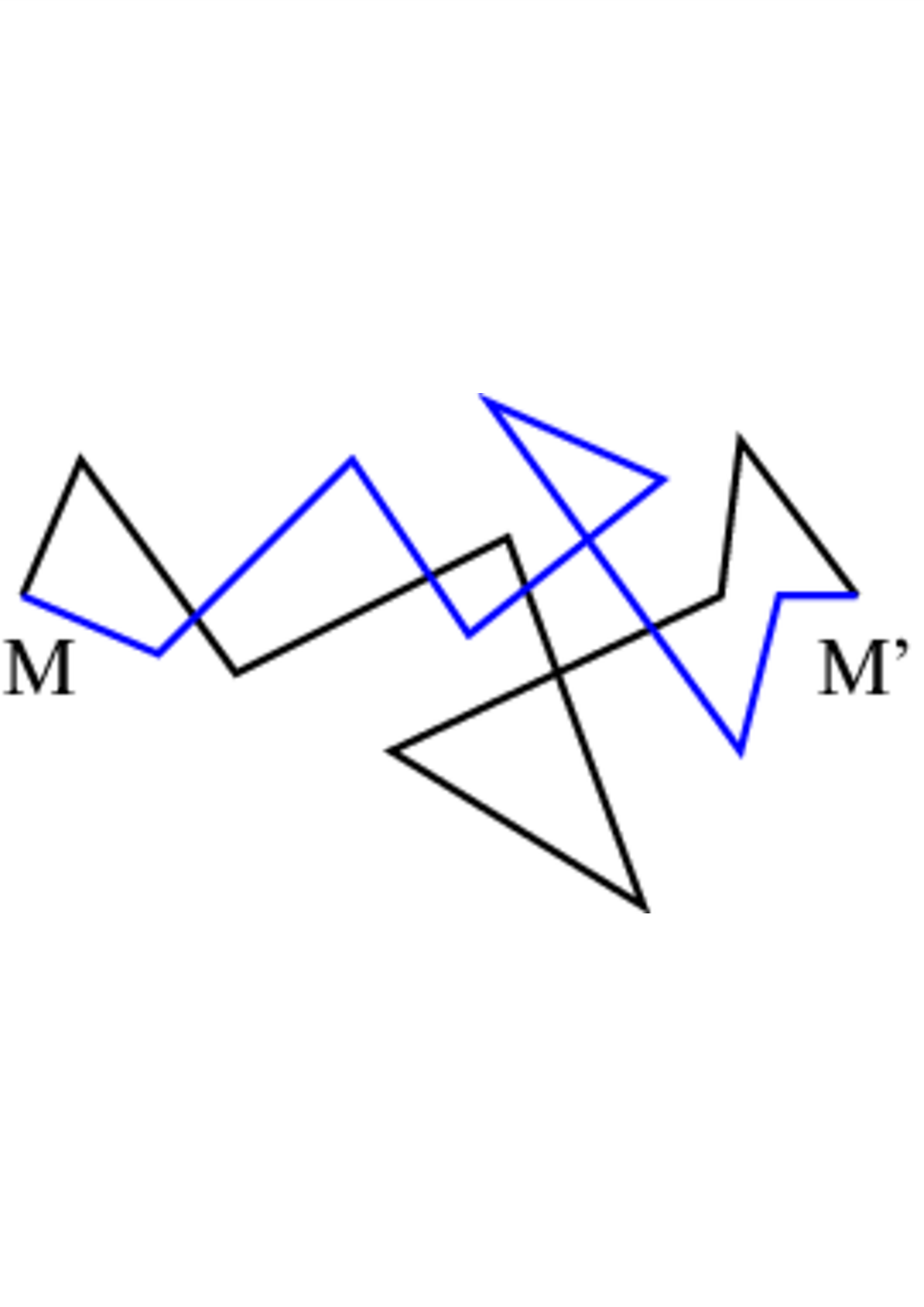}
    \caption{Two typical realizations of a monitored quantum walk on a graph from $|\br_{M'}\rangle$ to
    $|\br_M\rangle$ with $8$ steps between projective measurement,
    where the links represent elements of the evolution matrix ${\hat T}_M$.
    The measured state $|\br_M\rangle$ is only visited at the end of the quantum walk. The summation
    with respect to all possible quantum walk realizations gives the quantum walk sum (QWS) of 
    Eq. (\ref{qws1}).
    }    
\label{fig:qw}
\end{figure}

\section{Unitary vs. monitored evolution}
\label{sect:properties}

In the following we will calculate the transition amplitude for a quantum walk on a graph,
comprising $N$ sites $\{\br_j\}_{j=1,\ldots,N}$ with associated states $\{|\br_j\rangle\}$. 
These states can be assumed to be single particle states but any other finite-dimensional basis
is possible as well. The measured state is an element of the basis 
$|\psi\rangle=|\br_M\rangle$, and the initial state $|\psi_0\rangle=|\br_{M'}\rangle$ is
from the same basis. This choice will simplify the calculations but can also be generalized
in a straightforward manner. The definition of the graph does not provide a structure,
i.e., there is no distance. The latter will be created by the quantum walk. Then, for instance,
a distance between basis states can be defined by the transition probability between
different states. As we will see below, the resulting structure depends on the type of
quantum walk, e.g., whether it is based on a unitary or a monitored evolution. It also
depends on the eigenvectors of the evolution operator.  

Although we are mostly interested in the monitored evolution,
we consider the unitary evolution first. The transition amplitude reads in this case 
$\langle\br_M|e^{-iHt/\hbar}|\br_{M'}\rangle$
with a continuous time $t$ and the Hamiltonian $H$.
For the uniform state $|\psi_0\rangle=\sum_k|\br_k\rangle/\sqrt{N}$ we assume that it is invariant
under the unitary evolution: 
\beq
\label{groundstate}
e^{-iHt/\hbar}|\psi_0\rangle=|\psi_0\rangle
,
\eeq
such that this state is eigenstate of the evolution operator with
eigenvalue 1 and, consequently, an eigenstate of $H$ with zero eigenvalue. 
(In the following we will use the notation in which the Planck constant $\hbar$ is implicit 
in the definition of $H$.)
Assuming further that the spectrum of $H$ is non-negative, the zero energy energy eigenvector
means that the uniform state is a stationary groundstate. Moreover, Eq. (\ref{groundstate}) 
implies the detailed balance condition for the transition amplitude
\beq
\label{detailed_balance}
\sum_{k=1}^N\langle\br_j|e^{-iHt}|\br_{k}\rangle
=1
\ \ \  (t\ge0)
,
\eeq
which defines a uniform eigenvector with eigenvalues 1 for the evolution matrix
$(\langle\br_j|e^{-iHt}|\br_{k}\rangle)$, namely
$\bv_1=(1,\ldots,1)^T/\sqrt{N}$. The other eigenvectors of the unitary matrix 
$(\langle\br_j|e^{-iHt}|\br_{k}\rangle)$ must be orthogonal to $\bv_1$ for non-degenerate eigenvalues. 
It should be noted that for a classical random walk the detailed balance condition applies
to the transition probability, since there is no transition amplitude. For the unitary evolution
of the quantum walk the transition probability obeys detailed balance inherently:
$\sum_{k=1}^N\langle\br_j|e^{-iHt}|\br_{k}\rangle\langle\br_k|e^{iHt}|\br_{j}\rangle=1$.

For many practical questions regarding the unitary evolution it is convenient to average with
respect to the continuous time~\cite{2003ConPh..44..307K}.
An example is the transition probability. With eigenstates and eigenvalues of the Hamiltonian $H$,
$\{|E_k\rangle\}$ and $\{E_k\}$, we define $q_{j,k}=\langle\br_j|E_k\rangle$ and 
$q^*_{j,k}=\langle E_k|\br_j\rangle$, from which we get the orthonormality relation
$\sum_{k=1}^Nq^{}_{j,k}q_{j\rq{},k}^*=\delta_{jj'}$ and the time-average transition probability
\beq
\label{time_av}
\bar{P}_{kk\rq{}}
=\lim_{\epsilon\to0}\epsilon\int_0^\infty |\langle\br_k|e^{-iHt}|\br_{k\rq{}}\rangle|^2
e^{-\epsilon t}dt
=\sum_j |q_{k,j}|^2|q_{k^\prime,j}|^2
,
\eeq
where the second equation is valid
for non-degenerate eigenvalues. In the case of degenerate eigenvalues we must extend the summation 
over $j$ to a double summation over all pairs of degenerate states. 
It should be noted that this expression, as a result of the time average,
depends only on the spectral weights $\{q_{k,j}\}$ but not on the eigenvalues $\{E_k\}$.

As an alternative to the time-averaged unitary evolution,
we introduce repeated measurements which provides an evolution defined at discrete measurement
times~\cite{Gruenbaum2013}. Then our goal is to detect the state $|\br_M\rangle$ for the first time,
assuming that we only measure at discrete times $\tau,2\tau,\ldots$. In this case the measurement 
is performed by the projector $\Pi_M={\bf 1}-|\br_M\rangle\langle\br_M|$ with the identity operator 
${\bf 1}$. Thus, we obtain for the evolution of $|\br_{M\rq{}}\rangle$ a sequence 
of $m-1$ measurements with
$|\psi_m\rangle=(e^{-iH\tau}\Pi_M)^{m-1}e^{-iH\tau}|\br_{M\rq{}}\rangle$. Assuming 
that there exists an integer $m_f$ with $|\psi_{m_f}\rangle\propto|\br_M\rangle$, we get
$|\psi_m\rangle=0$ for all $m>m_f$. This means that the monitored evolution of $|\br_{M\rq{}}\rangle$ 
terminates at $m=m_f$, and that it is characterized by the finite sequence of transition amplitudes
$\phi_{MM\rq{}}(m,\tau)=\langle\br_M|(e^{-iH\tau}\Pi_M)^{m-1}e^{-iH\tau}|\br_{M\rq{}}\rangle$
for $1\le m\le m_f$.
The existence of $m_f<\infty$ depends on the details of the evolution~\cite{quancheng20}.
This was studied previously~\cite{Gruenbaum2013,Friedman2017} 
and will also be a subject of the present work.

The transition amplitude $\phi_{MM\rq{}}(m,\tau)$
can be rewritten as~\cite{PhysRevA.110.022208}
\beq
\label{trans_amp00}
\phi_{MM\rq{}}(m,\tau)
=\langle\br_M|e^{-iH\tau/2}T_M^{m-1} e^{-iH\tau/2}|\br_{M\rq{}}\rangle
\ , \ \
T_M:=e^{-iH\tau/2}\Pi_M e^{-iH\tau/2}
,
\eeq
where $T_M$ is the monitored evolution operator, the analogue to the unitary evolution operator
$e^{-iH\tau}$ for the monitored evolution. In contrast to the unitary evolution, the monitored evolution
does not obey detailed balance $\sum_{M'=1}^N|\Phi_{MM'}(m,\tau)|^2< 1$ for $m>1$, reflecting that the 
projective measurement effectively couple the quantum walk to the environment. Moreover, the sum of the
transition probabilities with respect to the number of measurements is restricted as
$\sum_{m\ge 1}|\Phi_{MM'}(m,\tau)|^2\le 1$. This is a consequence of the fact that
$\sum_{m'=1}^m|\Phi_{MM'}(m',\tau)|^2$ is the probability of the transition 
$|\br_{M'}\rangle\to|\br_M\rangle$ under the condition that $|\br_M\rangle$ was not measured for
any $2\le m'\le m$.
This monitoring approach was discussed in Ref.~\cite{Gruenbaum2013} and has been applied to single-particle states
to detect the particle location on a 
graph~\cite{Dhar_2015,dhar15,Friedman2017,lahiri19,PhysRevResearch.1.033086,PhysRevResearch.2.033113,
PhysRevResearch.5.023141} as well as to spin
systems~\cite{liu2024b}.
Its advantage, in comparison to a unitary evolution, is that 
all information about the transition $|\br_{M\rq{}}\rangle\to|\br_M\rangle$
is recorded in $m_f$ measurements. 

The amplitude for the monitored transition 
in Eq. (\ref{trans_amp00}) can be rewritten as a quantum walk sum (QWS) (cf. App. \ref{app:qws})
\beq
\label{qws1}
\phi_{MM'}(m,\tau)=\sum_{k_1,\ldots,k_m}
z_1^2z_2^2\cdots z_m^2
q_{M,k_1}K_{M;k_1,k_2}\cdots K_{M;k_{m-1},k_m}q^*_{M\rq{},k_m}
\eeq
with $z_k=e^{-iE_k\tau/2}$. Two realizations of a QWS are visualized in Fig. \ref{fig:qw}.
The matrix elements of the evolution matrix ${\hat T}_M$ and the kernel $K_M$ are related as
\beq
\label{energy_basis}
{\hat T}_{M;kl}:=
\langle E_k|T_M|E_l\rangle=z_kK_{M;kl}z_l
\ \ {\rm with}\ \ 
K_{M;kl}:=\delta_{kl}-q_{M,k}^*q^{}_{M,l}=\sum_{j\ne M}q_{j,k}^*q^{}_{j,l}
.
\eeq
The last equation is a consequence of the fact that $\{|\br_k\rangle\}$ is a complete basis. 
The kernel $K_M$ is a projector (i.e., $K_M^2=K_M$) and satisfies the eigenvalue equation
\beq
\label{ONS}
\sum_{l=1}^NK_{M;kl}q^*_{M',l}
=(1-\delta_{MM'})q_{M',k}^*
\eeq
with the eigenvalue $\nu_M=0$ for the eigenvector $\bQ_M^*=(q^*_{M,1},\ldots,q^*_{M,N})^T$
and $N-1$ degenerate eigenvalues $\nu_{M'}=1$ ($M'\ne M$) for the eigenvectors $\bQ_{M'}^*$.

Multiplication of the kernel from both sides 
with the diagonal unitary matrix $D^{1/2}={\rm diag}(z_1,\ldots,z_N)$
distributes the eigenvalues $\{\lambda_k\}$ 
of $T_M$ over the entire unit disk, leaving only the zero eigenvalue $\nu_M$ unchanged: 
$\lambda_M=\nu_M=0$ . As long as $|\lambda_k|<1$ the corresponding eigenstates decay 
exponentially under the evolution with ${\hat T}_M^{m-1}$. However, in the
presence of a degeneracy $z_k=z_l$ there exist also eigenvalues at the circular boundary of the 
disk with $|\lambda_l|=1$ (cf. App. \ref{app:degeneracy}), whose eigenstates do not 
decay~\cite{PhysRevResearch.2.033113,PhysRevResearch.5.023141}.
We will see in the next section that energy orthogonal states have a similar effect.

\section{Localized states}
\label{sect:loc_states}

The definition of a localized state $|\br_k\rangle$ requires a reference basis. 
In our specific case this is the energy eigenbasis $\{|E_j\rangle\}$ of the Hamiltonian $H$
and the unitary evolution operator $\exp(-iH\tau)$.
To understand the main features of localized states, we consider special realizations.
The extreme example of a localized basis $\{|\br_k\rangle\}$ is the energy eigenbasis itself.
In this case the measured state is also an energy eigenstate, and we get 
$q_{k,j}=\delta_{jk}$ for all $j,k$. Then Eq. (\ref{energy_basis}) is the diagonal evolution matrix
\beq
{\hat T}_{M;jk}=z_k^2\delta_{jk}(1-\delta_{Mk})
.
\eeq
From this result we get for Eq. (\ref{trans_amp00}) the transition amplitude 
$\phi_{MM'}(1,\tau)=\delta_{MM\rq{}}$ and $\phi_{MM'}(m,\tau)=0$ for any $m\ge 2$, 
which is identical with the quantum Zeno limit $\tau\sim0$ directly from 
Eq. (\ref{trans_amp00}).
This is a consequence of the unitary evolution between projective measurements
that changes only the phase of the initial state but does not allow any transition to other states.
This is also the case for the time-average transition probability in Eq. (\ref{time_av}),
which becomes the unit matrix in this basis.
In contrast, the extreme example for delocalized states are plane waves
\beq
\label{plane_wave_basis}
q_{k,j}=\frac{1}{\sqrt{N}}e^{2\pi i (k-1)(j-1)/N}
,
\eeq
which yields for the monitored evolution the matrix elements
\beq
\label{deloc_mon}
\langle E_k|T_M|E_l\rangle
=
z_k^2\delta_{kl}-\frac{1}{N}e^{-2\pi i (k-l)(M-1)/N}z_kz_l
.
\eeq
When the initial state is localized to some subspace, a unitary evolution will not leave this
localized space such that all states outside the localized space cannot be reached. 
Next it will be shown that this is different in the case of the monitored evolution. 

First, we clarify what is formally understood as localization in the context of quantum walks 
on a finite-dimensional Hilbert space. We define a state $|\br_k\rangle$ as localized in terms of
the energy eigenstates if the inverse participation ratio
\beq
\label{loc_criterion}
c_k:=\sum_{j=1}^N|q_{k,j}|^4
\eeq
does not vanish asymptotically for $N\sim\infty$. 
(Note that $c_k:=\sum_{j=1}^N|q_{k,j}|^4\ne\sum_{k=1}^N|q_{k,j}|^4$.
The latter was used in Ref. \cite{Ziegler_2024}.)
In contrast, $|\br_k\rangle$ is delocalized when 
$c_k\sim 1/N$ for $N\sim\infty$. Thus, a localized state $|\br_k\rangle$ is defined by the
Hamiltonian $H$ or the unitary evolution operator that has 
an overlap with only a few energy eigenstates $|E_j\rangle$. More specific, the number of 
overlapping states remains finite when we increase the size of the Hilbert space.
While for the extremely localized example of $q_{k,j}=\delta_{jk}$ the inverse participation ratio is
$c_k=1$ ($1\le j\le N$), for the plane-wave basis the inverse participation ratio is $c_k=1/N$ for 
$1\le k\le N$.

We note that the concept of localization can also be formulated for the unitary evolution in 
terms of the time averaged transition probability of Eq. (\ref{time_av}).
It refers to the correlation of two states during the evolution, namely $|\br_{k}\rangle$ 
and $|\br_{k'}\rangle$, and is characterized as an exponential decay with $|\br_k-\br_{k'}|$. 
This definition is more common in the context of quantum transport and will not be used here. 
It should be noted that the inverse participation ratio $c_k$ is the time-averaged return probability 
$\bar{P}_{kk}$ of the unitary evolution. The fact that $\bar{P}_{kk}\sim N^0$ for localized
states reflects that the quantum walk returns frequently to the initial state upon time average,
while $\bar{P}_{kk}\sim 1/N$ for delocalized states indicates that the return of the quantum walk
to the initial state vanishes asymptotically for large systems.

The basis $\{|\br_k\rangle\}$ may have a state $|\br_l\rangle$ for a fixed $l$ that is orthogonal to the 
energy eigenstate $|E_j\rangle$ with fixed $j$: $q_{l,j}=\langle\br_l|E_j\rangle=0$. 
Such a state $|\br_l\rangle$ will be called subsequently an energy orthogonal state (EOS).
An EOS corresponds to the zero charge in the charge picture of Ref.~\cite{quancheng20},
where the charge is $p_j=|q_{M,j}|^2$.
A localized state can be constructed from a set of EOSs, which obeys $c_k\sim N^0$.
Then Eq. (\ref{energy_basis}) implies with $q_{M,j}=0$ for a fixed $j$ an isolated diagonal element 
for the monitored evolution matrix as
\beq
\label{local1}
{\hat T}_{M;jj'}={\hat T}_{M;j'j}=z^2_j \delta_{jj'}
\eeq
for all $j'$ and with the eigenvalue $z_j^2$ on the unit circle. 
This matrix element prevents transitions
$|E_j\rangle\to|E_{j'}\rangle$ as well as  transitions $|E_{j'}\rangle\to|E_j\rangle$,
which isolates the energy eigenstate $|E_j\rangle$ from the evolution of the other energy eigenstates.
More general, EOSs with $q_{M,j}=0$ for $j\in{\cal L}_M$ imply
\beq
\label{reduction1}
{\hat T}_{M;jj'}
=\cases{
z_jz_{j'}(\delta_{jj'}-q^*_{M,j}q^{}_{M,j'}) & for $j,j'\ne{\cal L}_M$ \cr
z_j^2\delta_{jj'} & otherwise \cr
}
,
\eeq
which yields with $q_{M',j}=0$ for $j\in{\cal L}_{M\rq{}}$
the monitored transition amplitude
\beq
\label{qws_loc}
\phi_{MM\rq{}}(m,\tau)=\sum_{k\notin{\cal L}_M,l\notin{\cal L}_M\cup{\cal L}_{M'}}
q_{M,k}z_k[{\hat T}_M^{m-1}]_{kl}q_{M',l}^*z_l
\]
\[
=\sum_{j_1,\ldots,j_{m-1}\notin{\cal L}_M,j_m\notin{\cal L}_M\cup{\cal L}_{M'}}
z_{j_1}^2z_{j_2}^2\cdots z_{j_m}^2q_{M,j_1}
(\delta_{j_1j_2}-q^*_{M,j_1}q^{}_{M,j_2})\cdots(\delta_{j_{m-1}j_m}-q^*_{M,j_{m-1}}q^{}_{M,j_m})
q^*_{M\rq{},j_m}
.
\eeq
This result corresponds with the QWS in Eq. (\ref{trans_amp01a}) of App. \ref{app:qws} when
$\sum_{k_m\ne M} q^*_{k_m,j_{m-1}}q_{k_m,j_m}$ is replaced by 
$\delta_{j_{m-1}j_m}-q^*_{M,j_{m-1}}q^{}_{M,j_m}$ due to the completeness relation
in Eq. (\ref{energy_basis}).
Thus, only the elements $q_{M,j}$ and $q^*_{M\rq{},j}$ ($j=1,\ldots, N$) appear in the
QWS, and $q_{M,j}=0$ for $j\in{\cal L}_M$ reduces the summation further due to Eq. (\ref{reduction1}).
This means that we can replace ${\hat T}_M$ by
the projected matrix $\Pi_{{\cal L}_M}{\hat T}_M\Pi_{{\cal L}_M}$,
where $\Pi_{{\cal L}_M}$ projects onto the space with $j\notin{\cal L}_M$,
in the QWS:
\beq
\label{qws_loc2}
\phi_{MM\rq{}}(m,\tau)=\sum_{k\notin{\cal L}_M,l\notin{\cal L}_M\cup{\cal L}_{M'}}
q_{M,k}z_k[\Pi_{{\cal L}_M}{\hat T}_M\Pi_{{\cal L}_M}]^{m-1}_{kl}q_{M',l}^*z_l
.
\eeq
This transition amplitude vanishes if $q_{M,j}q_{M',j}^*=0$ for all $j$. Moreover,
since the projection $\Pi_{{\cal L}_M}$ restricts the evolution to a subspace of the
$N$--dimensional Hilbert space, the number of independent quantum walks is also
restricted. Formally this means that we have the same evolution for
different initial states $|\br_{M'}\rangle$, $|\br_{M''}\rangle$ ($M'\ne M''$)
when the relation 
\beq
\label{relation2}
{\bf Q}_M\cdot\Pi_{{\cal L}_M}{\bf Q}_{M'}^*
={\bf Q}_M\cdot\Pi_{{\cal L}_M}{\bf Q}_{M''}^*
\eeq
is valid.

\begin{figure}
a)
\includegraphics[width=7.8cm,height=7.4cm]{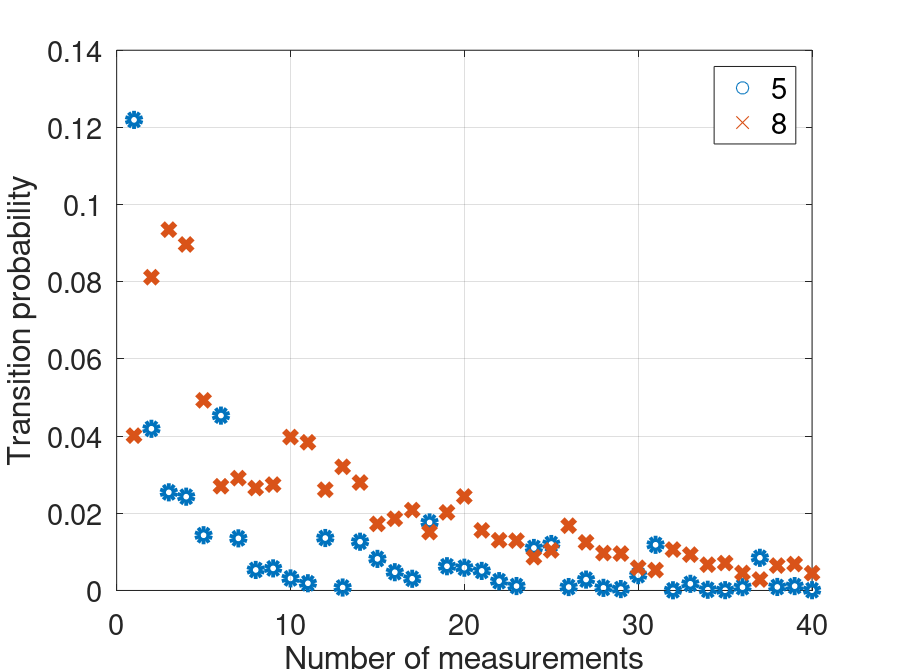}
b)
\includegraphics[width=7.8cm,height=7.4cm]{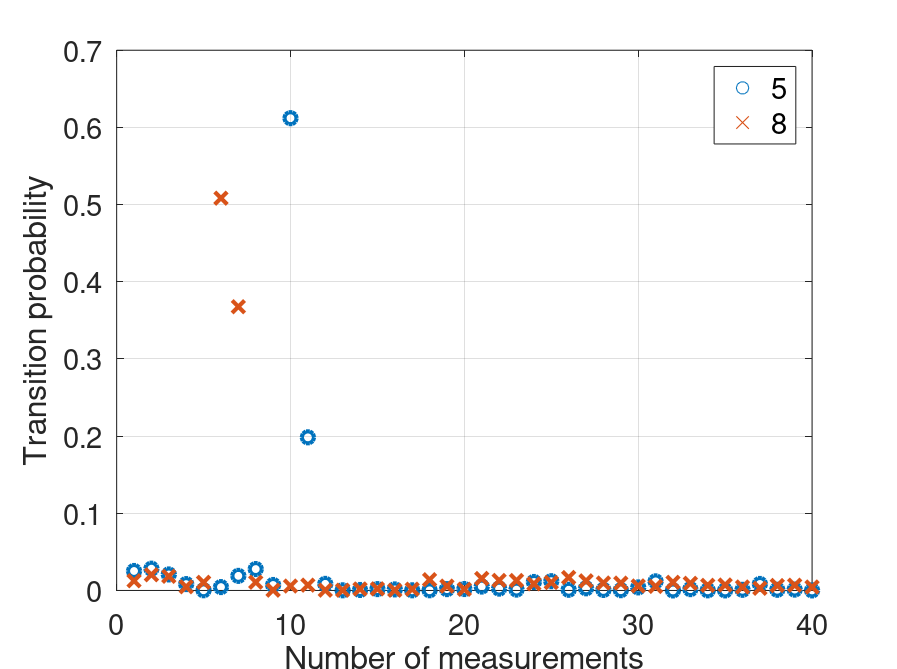}
    \caption{Transition probabilities for $|\br_{j}\rangle\to|\br_5\rangle$ 
    with $J\tau=\hbar$ as a function of the number of measurements
    a) in the localized basis and b) in the plane-wave basis.
           }    
\label{fig:tprob}
\end{figure}

There are several physical quantities which can be derived from the transition amplitude $\Phi_{MM'}(m,\tau)$.
Here we will focus on the transition probability $|\Phi_{MM'}(m,\tau)|^2$ and the mean transition time
\beq
\label{mfdt00}
{\bar\tau}_{{\cal N};MM'}:=\tau\frac{\sum_{m= 1}^{\cal N} (m-1)|\Phi_{MM\rq{}}(m,\tau)|^2}
{\sum_{m= 1}^{\cal N}|\Phi_{MM\rq{}}(m,\tau)|^2}
\eeq
for at most ${\cal N}-1$ measurements.
In general, the transition probability is a strongly fluctuating quantity with 
$0\le |\Phi_{MM'}(m,\tau)|^2\le 1$ (cf. Fig. \ref{fig:tprob}).
However, for $M'=M$ its sum is restricted as $\sum_{m\ge 1}|\Phi_{MM}(m,\tau)|^2=1$ for any (random) 
sequence of projective measurements if $m_f<\infty$, as shown in Ref. \cite{Ziegler_2021}. 
This means that the return probability is always 1 for sufficiently many measurements. On the other hand,
for $M\ne M'$ we have only $\sum_{m\ge 1}|\Phi_{MM'}(m,\tau)|^2\le 1$, where the actual value depends
on the details of the quantum walk. For instance, in the quantum Zeno limit the
probability even vanishes for the  transition $|\br_{M\rq{}}\rangle\to|\br_M\rangle$ ($M'\ne M$).
Thus, the quantum walk is characterized by the matrix structure of $|\Phi_{MM'}(m,\tau)|^2$ 
with respect to $M$ and $M'$, as well as by the mean transition time $\tau_{MM'}$ for
$|\br_{M\rq{}}\rangle\to|\br_M\rangle$.
This is also known as the mean first detected transition (FDT) time~\cite{quancheng20}.

\section{Recursive properties of the monitored evolution}
\label{sect:recursion}

A fundamental property of the unitary evolution of a closed quantum system, 
in contrast to classical random walks, is its non-stationary behavior~\cite{2003ConPh..44..307K}.
The question is whether or not monitoring by repeated measurements creates stationary states?
We will discuss below that there is not a unique answer to this question.

The monitored evolution is determined by the complex eigenvalues $\{\lambda_k\}$ of $T_M$,
which depend on $\{q_{M,k}\}$ as well as on $\{E_k\}$. In particular,
the decay of the evolution after $m-1$ measurements is proportional to the largest $|\lambda_k|^{m-1}$.
As already mentioned in Sect. \ref{sect:properties}, the eigenvalues of $T_M$ are distributed over the 
complex unit disk. These eigenvalues are quite sensitive to a change of $E_j\tau$. As long as all
eigenvalues are inside the unit disk (i.e., $|\lambda_k|<1$), the transition amplitude vanishes
for all $m>m_f$ with $m_f<\infty$. This means that the stationary states are zero.

The situation is different when we have eigenvalues with $|\lambda_k|=1$.
Without using the eigenvalues of $T_M$, we can employ the recursion relation of the transition amplitude
(cf. App. \ref{app:recursion})
\beq
\label{recursion2}
\phi_{MM\rq{}}(m,\tau)=\sum_{k\ne M}
\langle \br_M|e^{-iH\tau}|\br_k\rangle\phi_{kM\rq{}}(m-1,\tau)
\eeq
with $\phi_{kM\rq{}}(m-1,\tau)
=\langle \br_k|[e^{-iH\tau}({\bf 1}-|\br_M\rangle\langle\br_M|)]^{m-2}e^{-iH\tau}|\br_{M\rq{}}
\rangle$
and with the initial expression for $m=1$
$
\phi_{MM\rq{}}(1,\tau)=\langle \br_k|e^{-iH\tau}|\br_{M\rq{}}\rangle
$.
An important question is whether or not there is a stationary solution ${\bar\phi}_{lk}$ of the recursion
for $m\sim\infty$? There is always ${\bar\phi}_{lk}=0$, which corresponds with a decaying
solution. However, as shown in App. \ref{app:stationary}, a nonzero solution exists when there is a $k$ 
with (i) an EOS $q_{M,k}=0$ and (ii) $z_k^2=1$. Condition (i) implies 
${\hat T}_{M;kl}={\hat T}_{M;lk}=\delta_{lk}$ for all $l$, which corresponds with 
Eq. (\ref{local1}). Condition (ii) means the restriction of the energy eigenvalue to 
$E_k\tau=0\ ({\rm mod}\ 2\pi)$. For other energy values the application of ${\hat T}_M$ leads to a
rotation $z_k^{m-1}$ on the complex unit circle. This is not stationary but represents a limit cycle.
Thus, non-vanishing stationary solutions exist but they are isolated in the evolution.

\begin{figure}[t]
    a)
 \includegraphics[width=8cm,height=6cm]{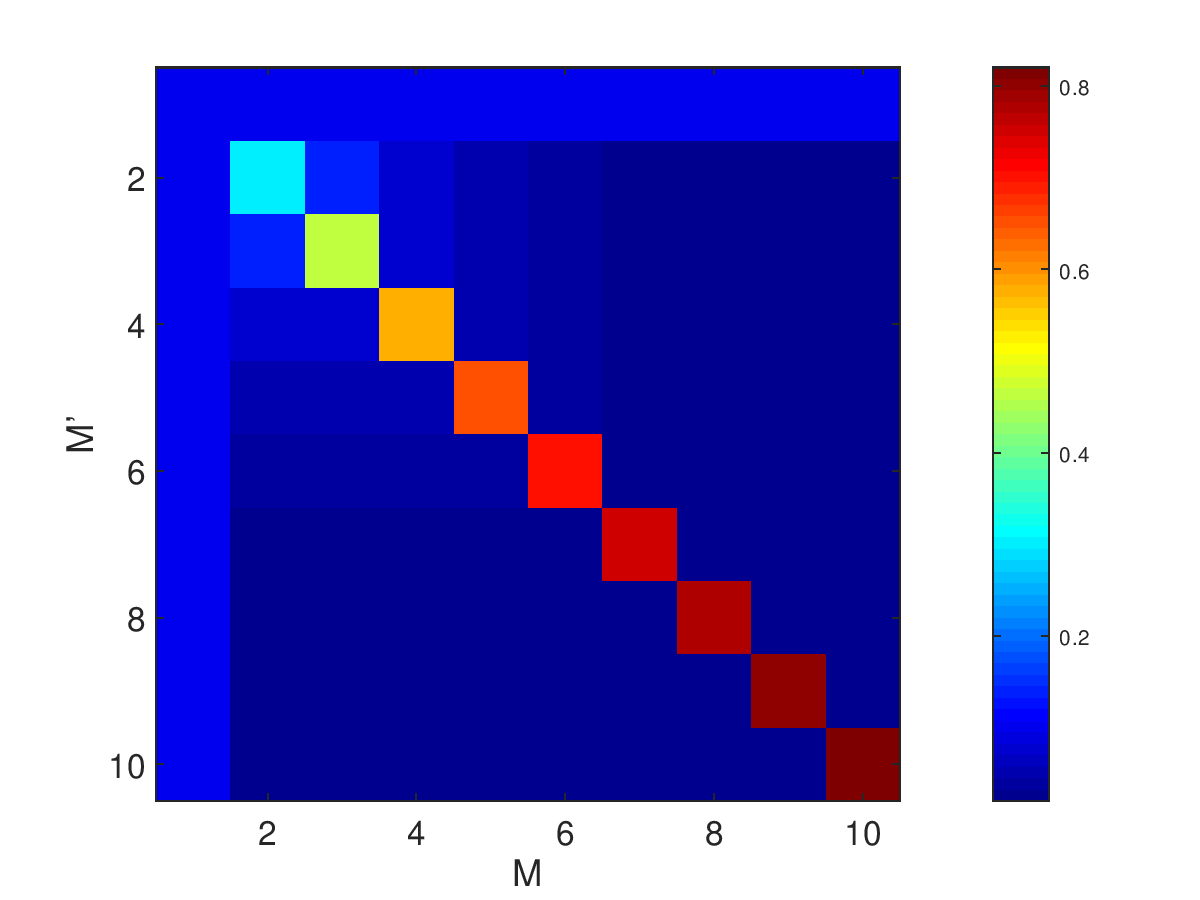}
 b)
 \includegraphics[width=8cm,height=6cm]{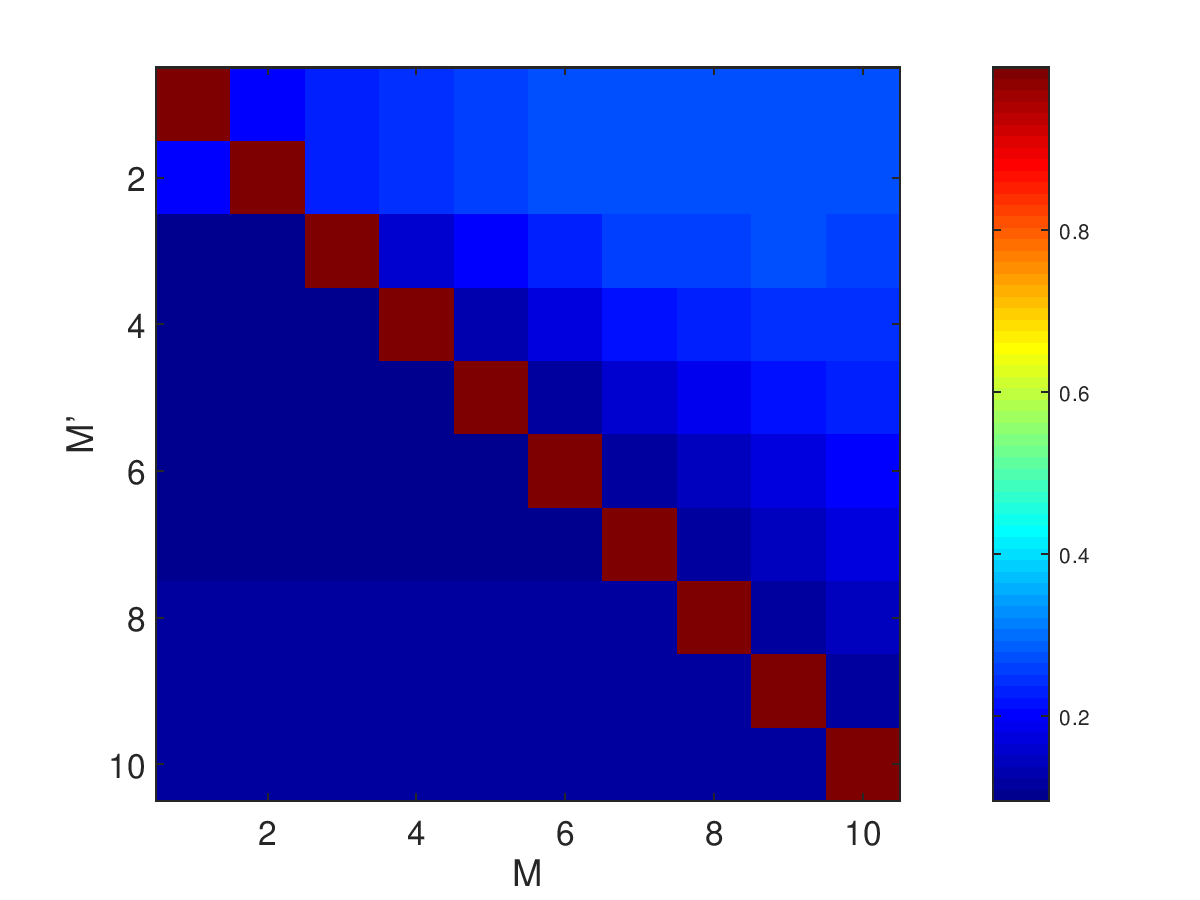}
 c)
\includegraphics[width=8cm,height=6cm]{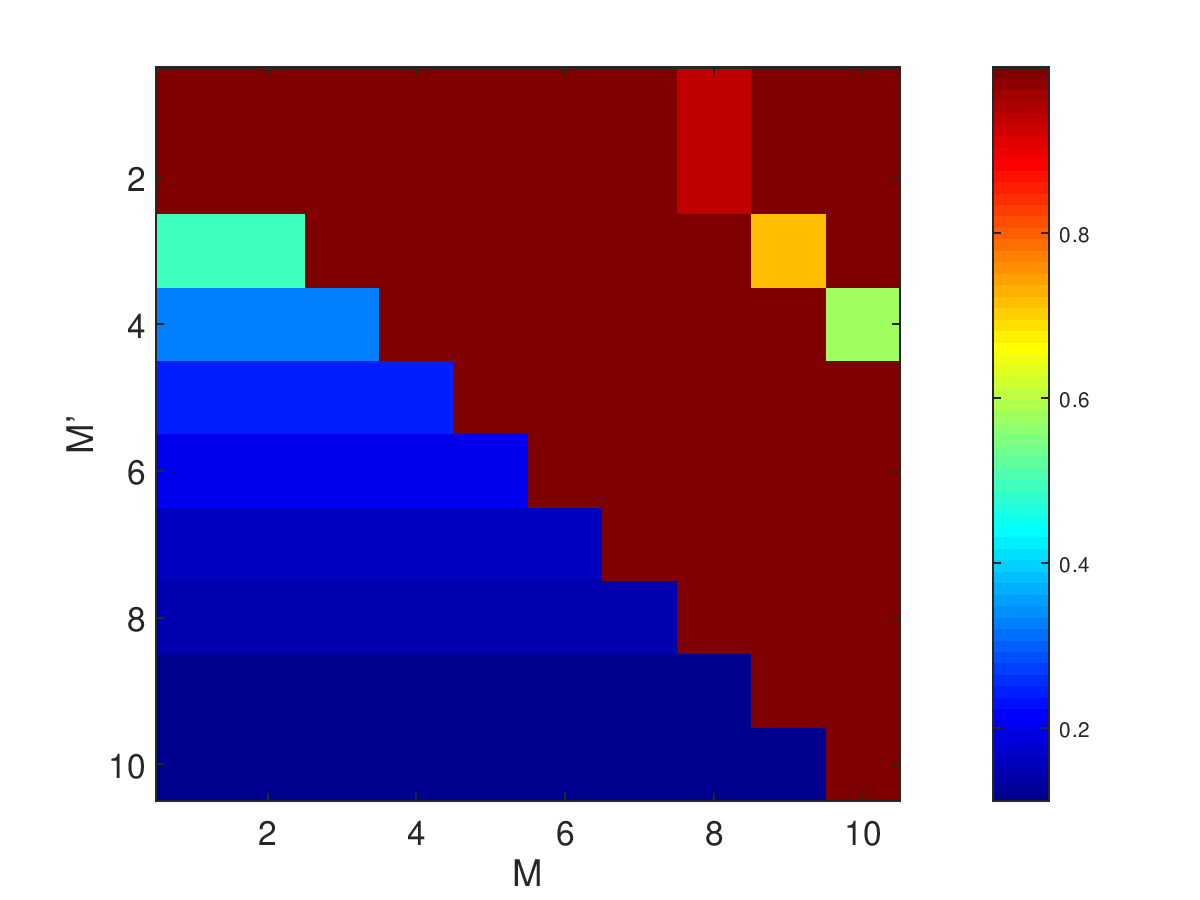}
 d)
\includegraphics[width=8cm,height=6cm]{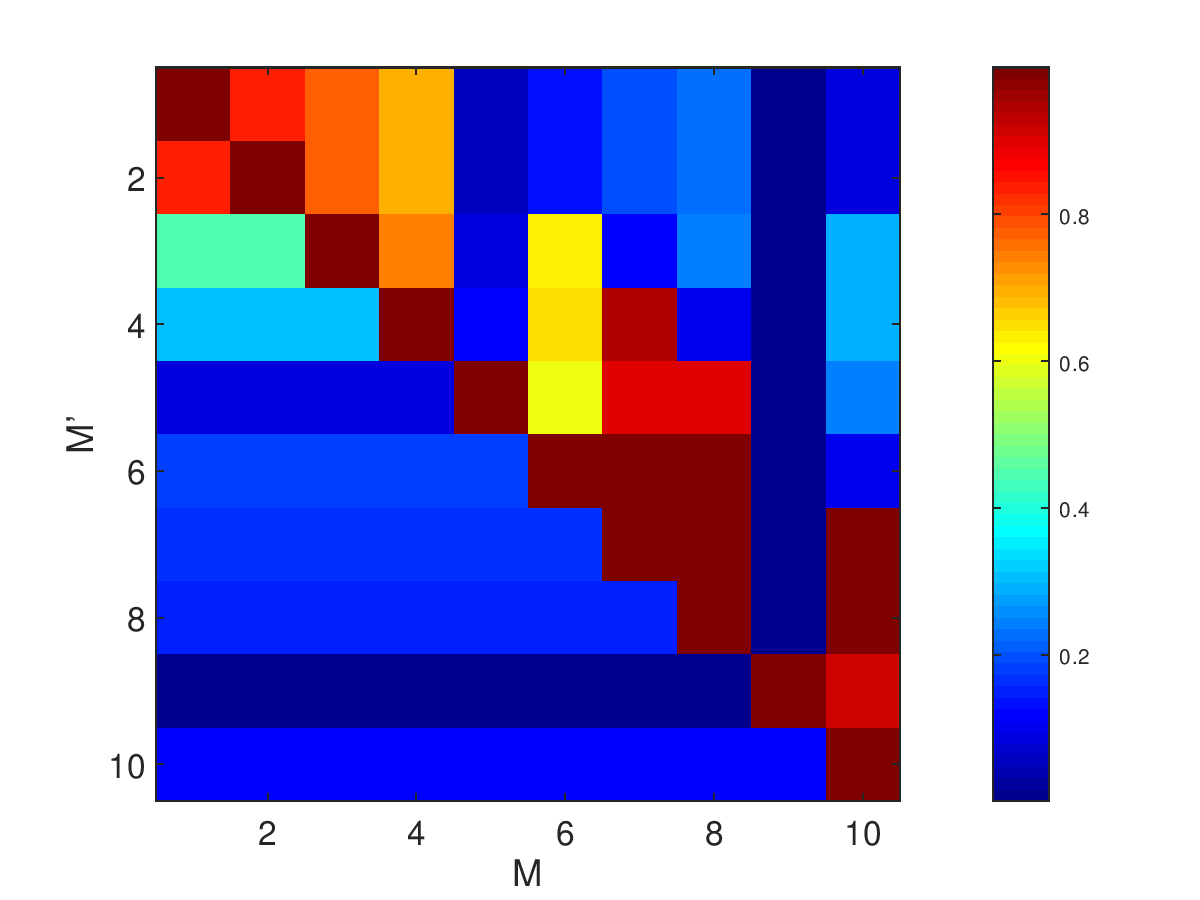}
    \caption{Transition probabilities on a graph with $N=10$ sites
    in the localized basis of Eq. (\ref{loc_basis2}).
    The calculation was performed for the linear spectrum $E_k=J(k-1)$ ($k=1,\ldots, N$). 
    This is also the case in the other examples of Figs. \ref{fig:tprob} -- \ref{fig:mfdt3}.
    a) The unitary evolution of the time-averaged transition probability
    ${\bar P}_{MM'}$ of Eq. (\ref{ue_transition}) indicates a typical
    localization effect. 
    b) For the monitored evolution the sum $\sum_{m=1}^{500}|\phi_{MM'}(m,\tau)|^2$ 
    reflects the quantum Zeno effect in b) for $J\tau=0.04\hbar$, 
    while in c) for $J\tau=\hbar$ and in d) for $J\tau=\pi\hbar/2$ an asymmetric transition
    is visible.
    }    
\label{fig:decay}
\end{figure}

\begin{figure}[t]
 \centering
a)
\includegraphics[width=5.8cm,height=5.6cm]{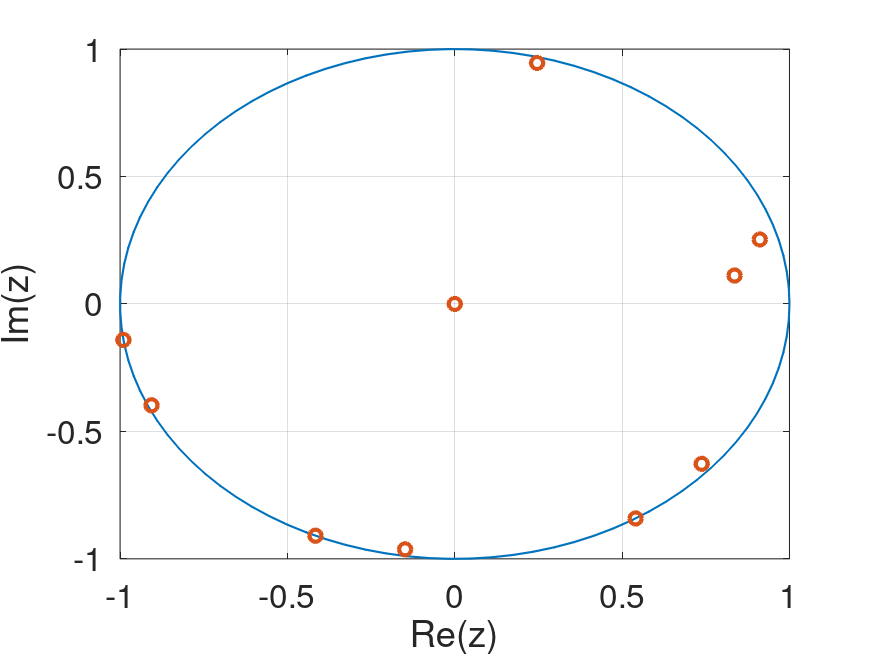}
b)
\includegraphics[width=5.8cm,height=5.6cm]{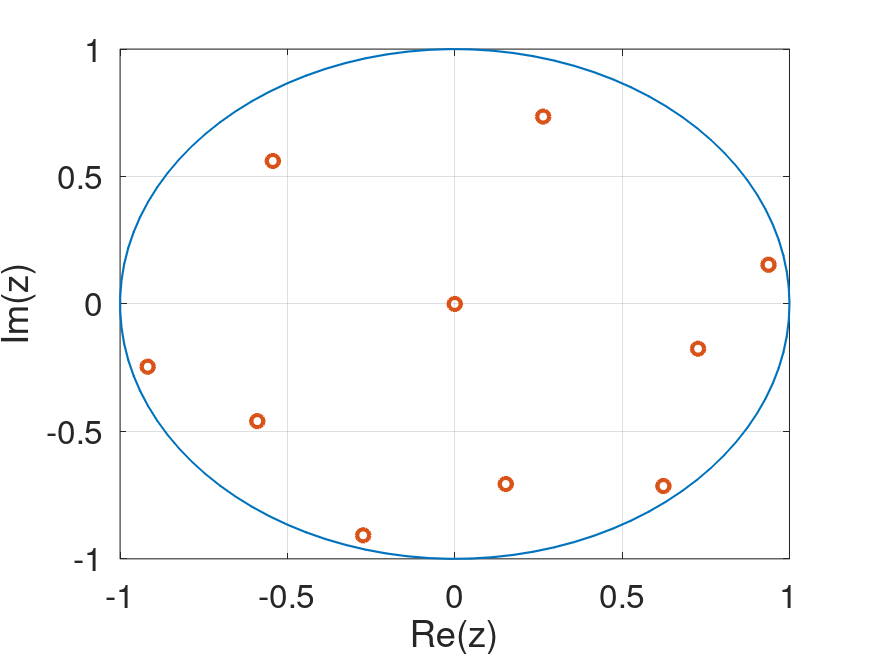}\\
c)
\includegraphics[width=7.5cm,height=6.6cm]{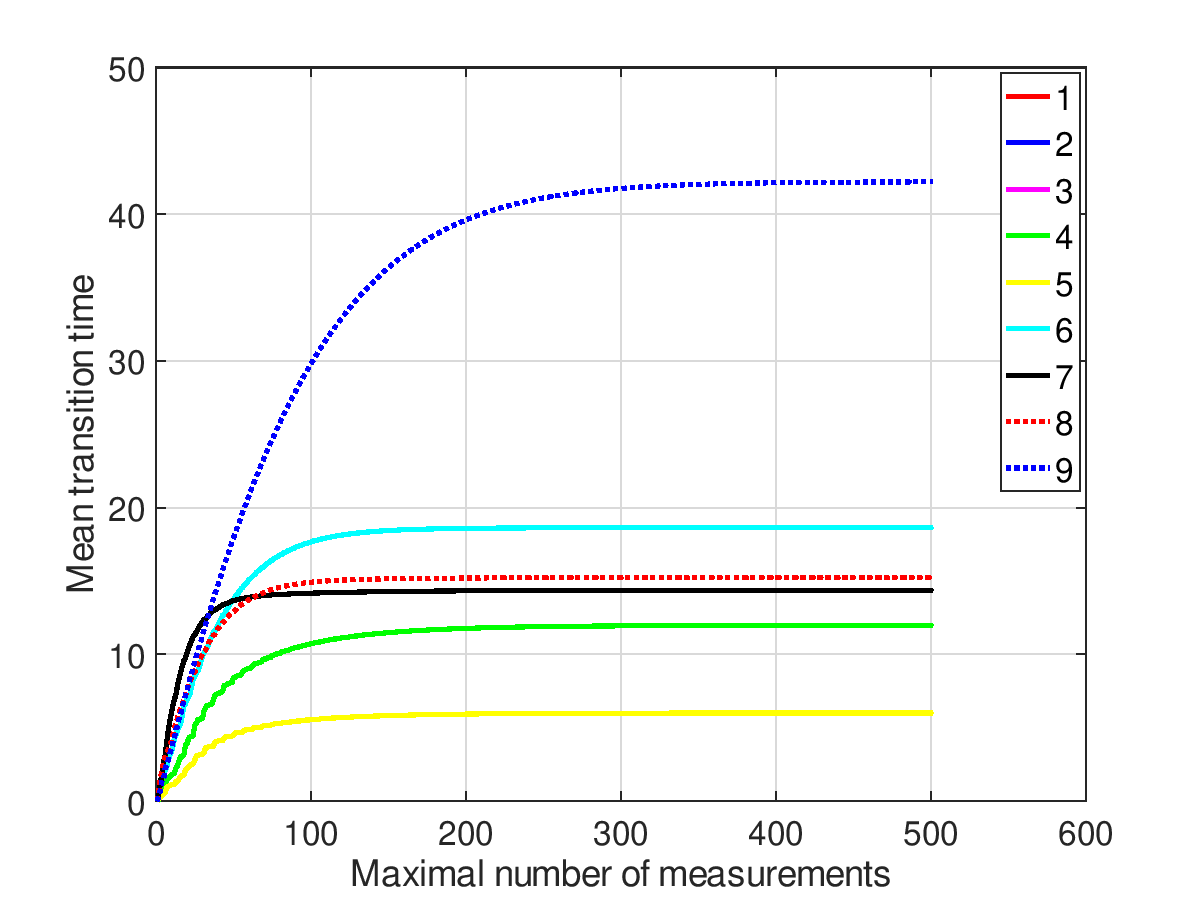}
d)
\includegraphics[width=7.5cm,height=6.6cm]{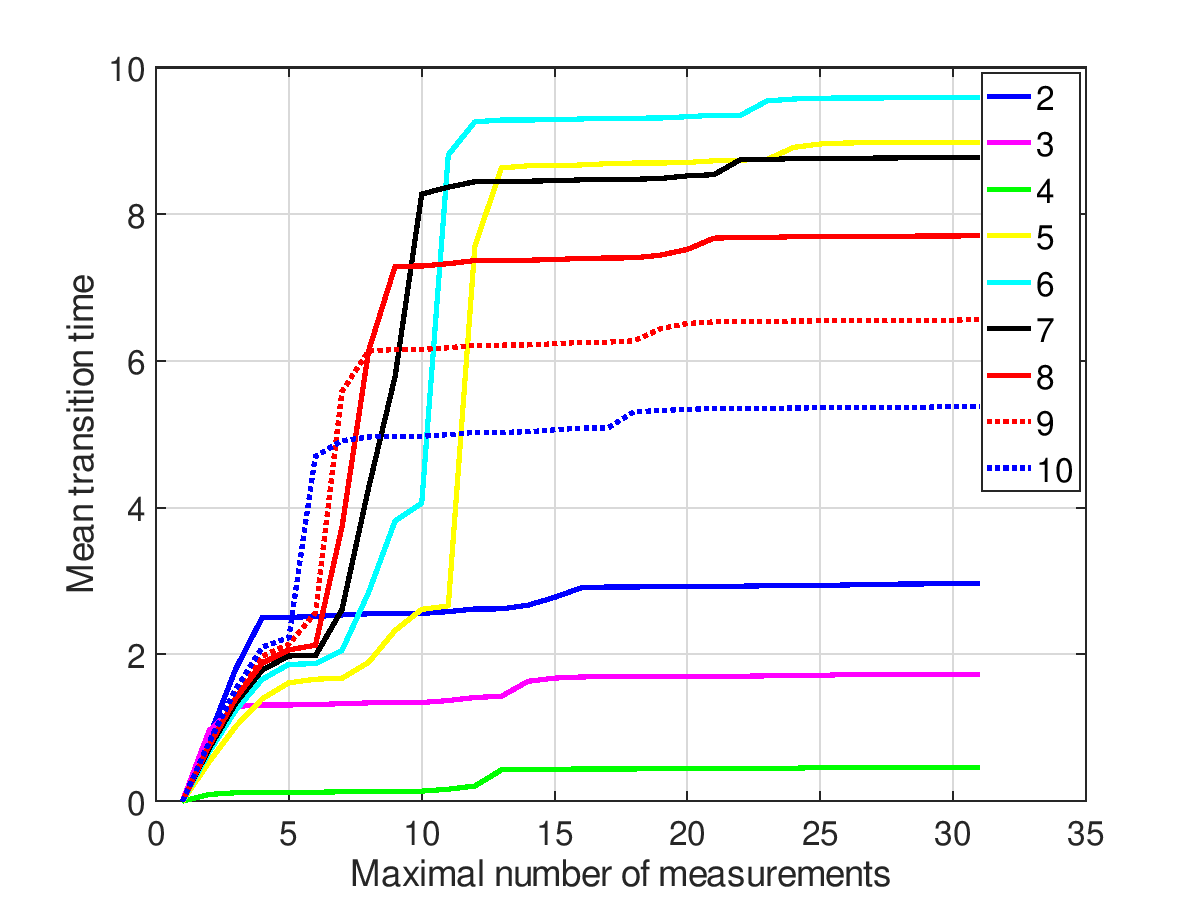}
    \caption{Eigenvalues of the monitored evolution matrix ${\hat T}_5$
    for a) the localized basis and b) the plane-wave basis with $J\tau=\hbar$.
    The mean transition times for ${\hat T}_5$ are plotted
    c) in the localized basis, where $j=1,2,3,4$ are equal, and d) in the plane-wave basis.
        }    
\label{fig:mfdt1}
\end{figure}

\begin{figure}[t]
 \centering
a)
\includegraphics[width=5.8cm,height=5.6cm]{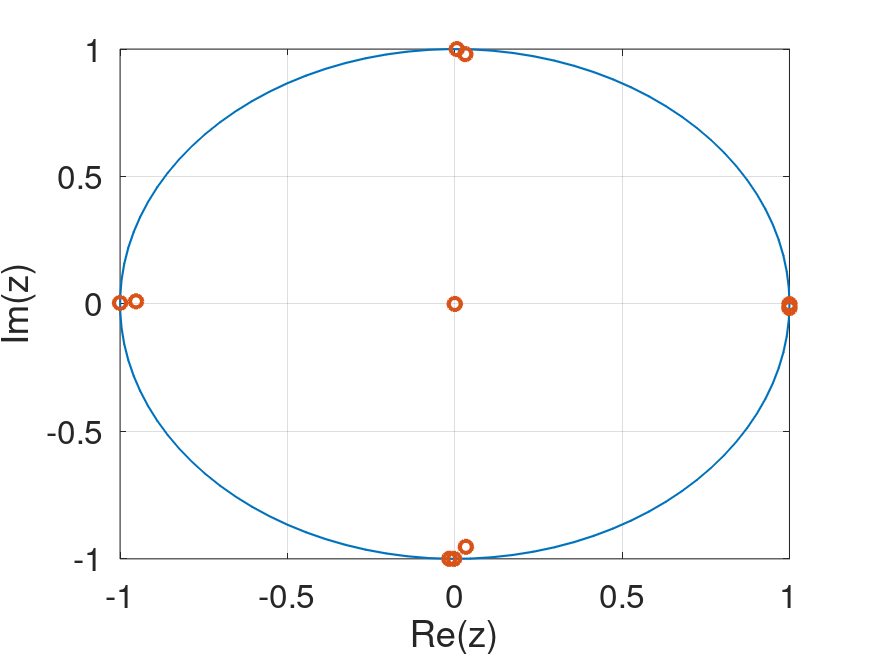}
b)
\includegraphics[width=5.8cm,height=5.6cm]{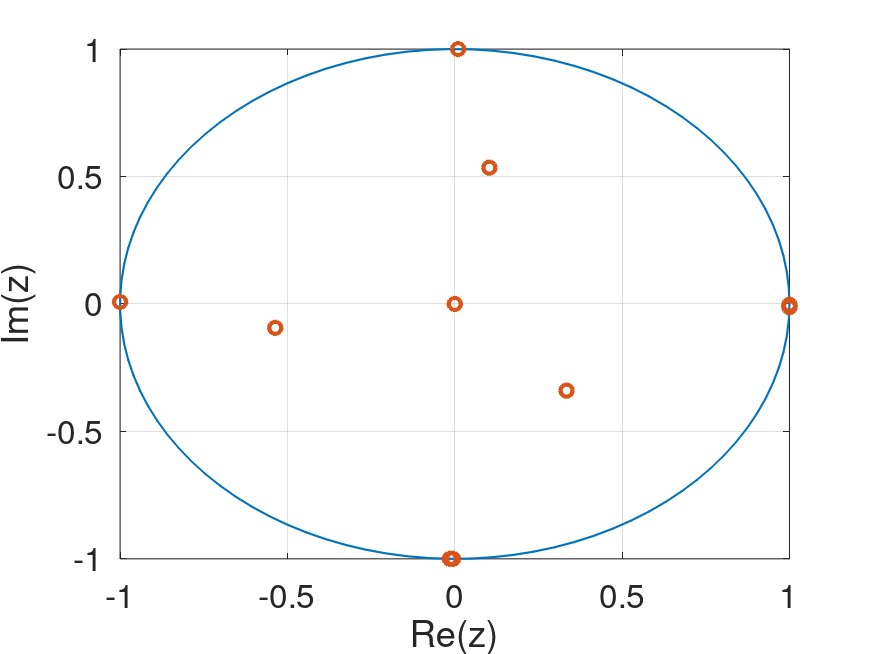}\\
c)
\includegraphics[width=7.5cm,height=6.6cm]{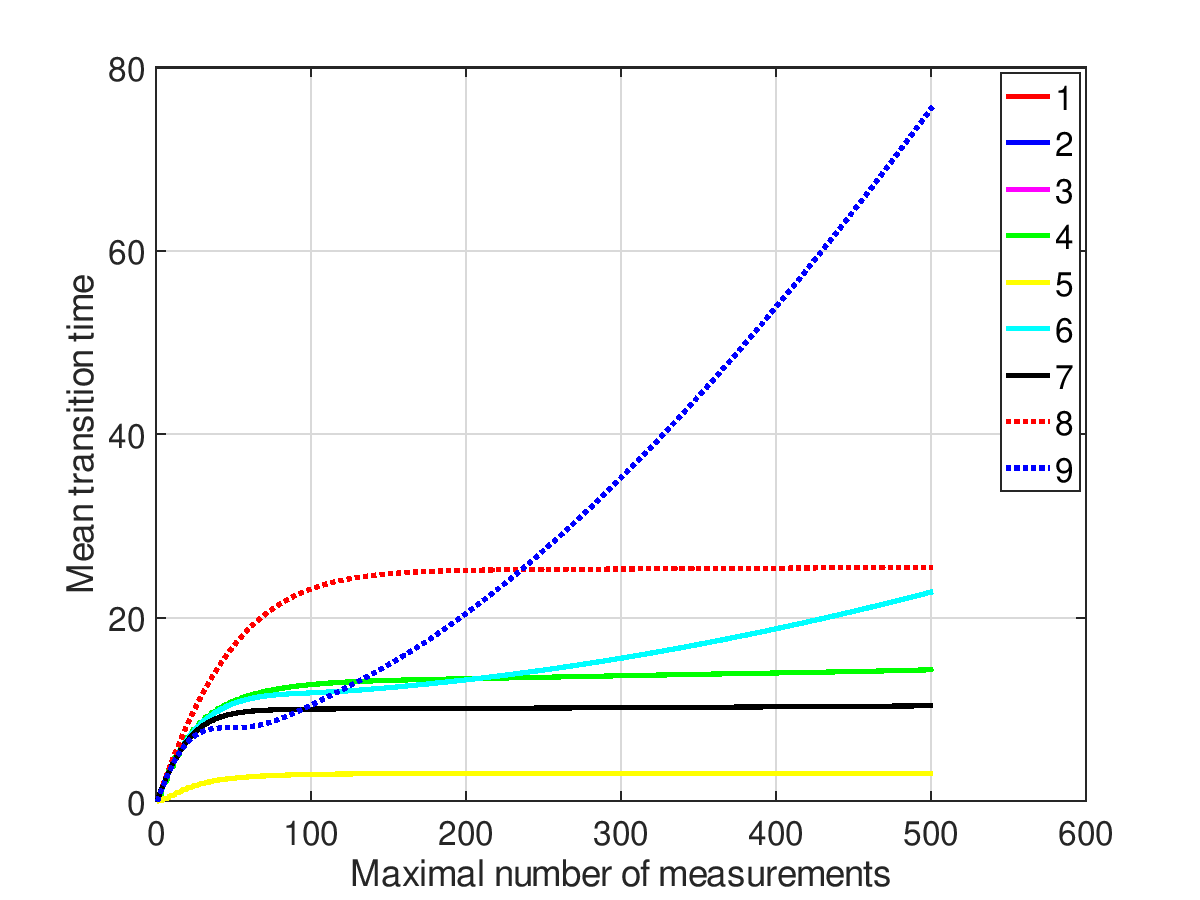}
d)
\includegraphics[width=7.5cm,height=6.6cm]{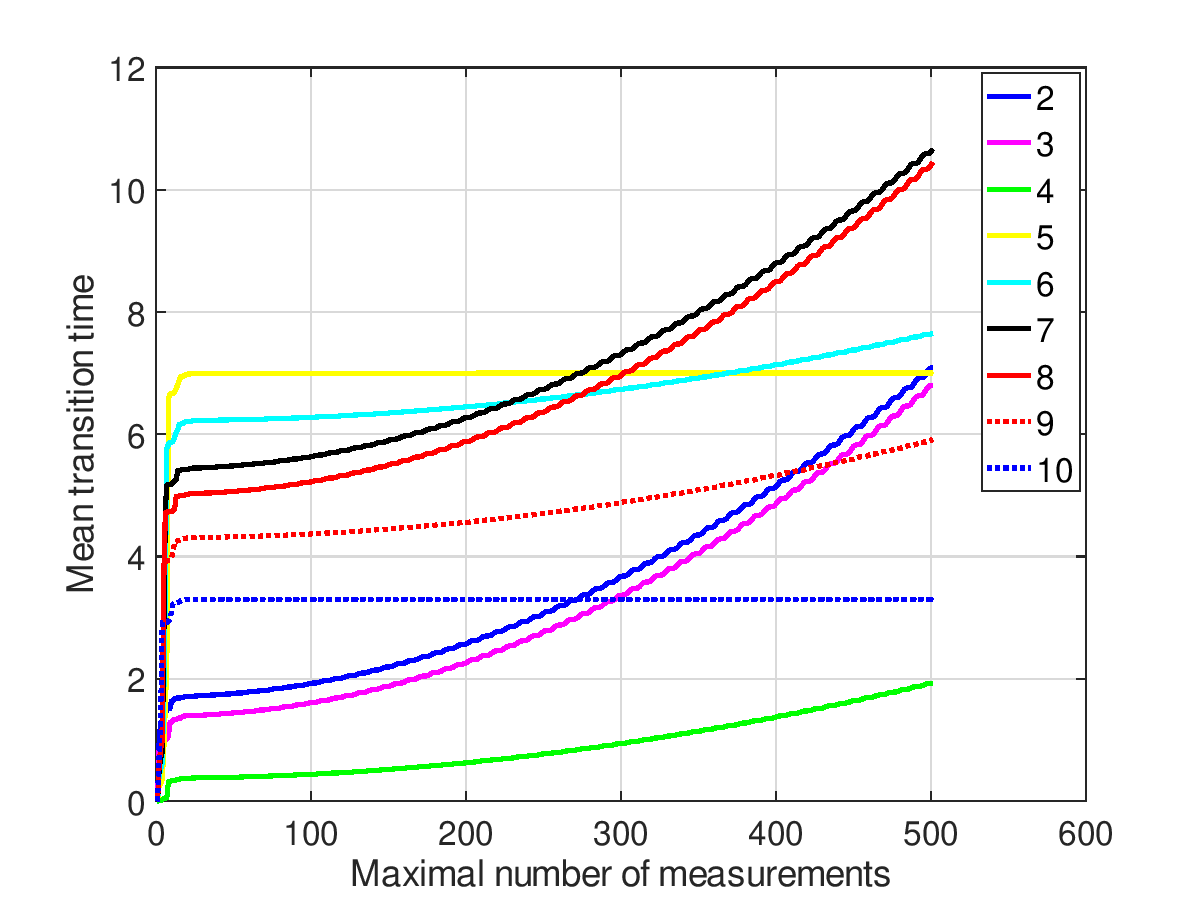}
    \caption{Eigenvalues of the monitored evolution matrix ${\hat T}_5$
    with $J\tau=(\pi/2+0.002)\hbar$ 
    a) in the localized basis and b) in the plane-wave basis.
    The mean transition times $|\br_j\rangle\to|\br_5\rangle$ are plotted
     c) in the localized basis and d) in the plane-wave basis.
    They diverge for some transitions
    due to the vicinity of spectral degeneracies. 
    }    
\label{fig:mfdt2}
\end{figure}

\begin{figure}[t]
a)
\includegraphics[width=7.5cm,height=6.6cm]{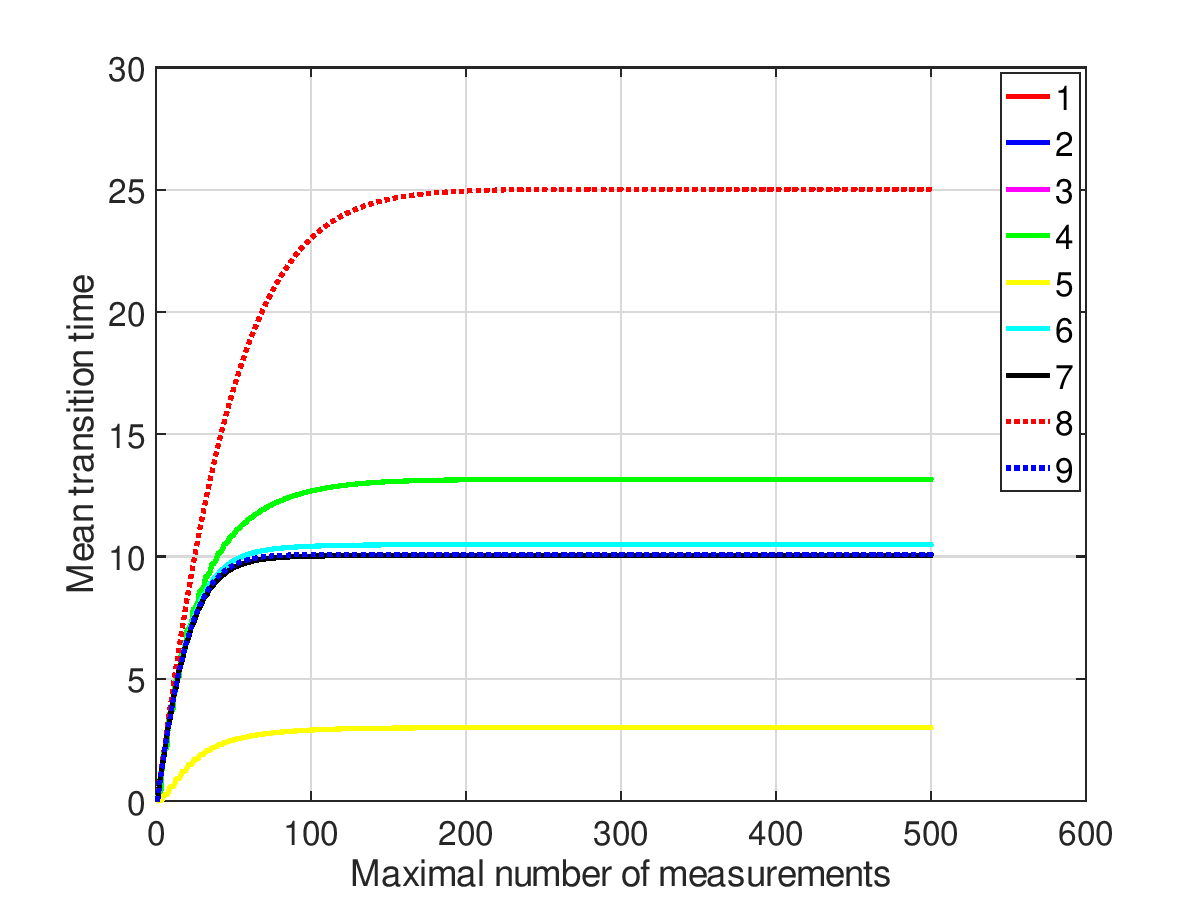}
b)
\includegraphics[width=7.5cm,height=6.6cm]{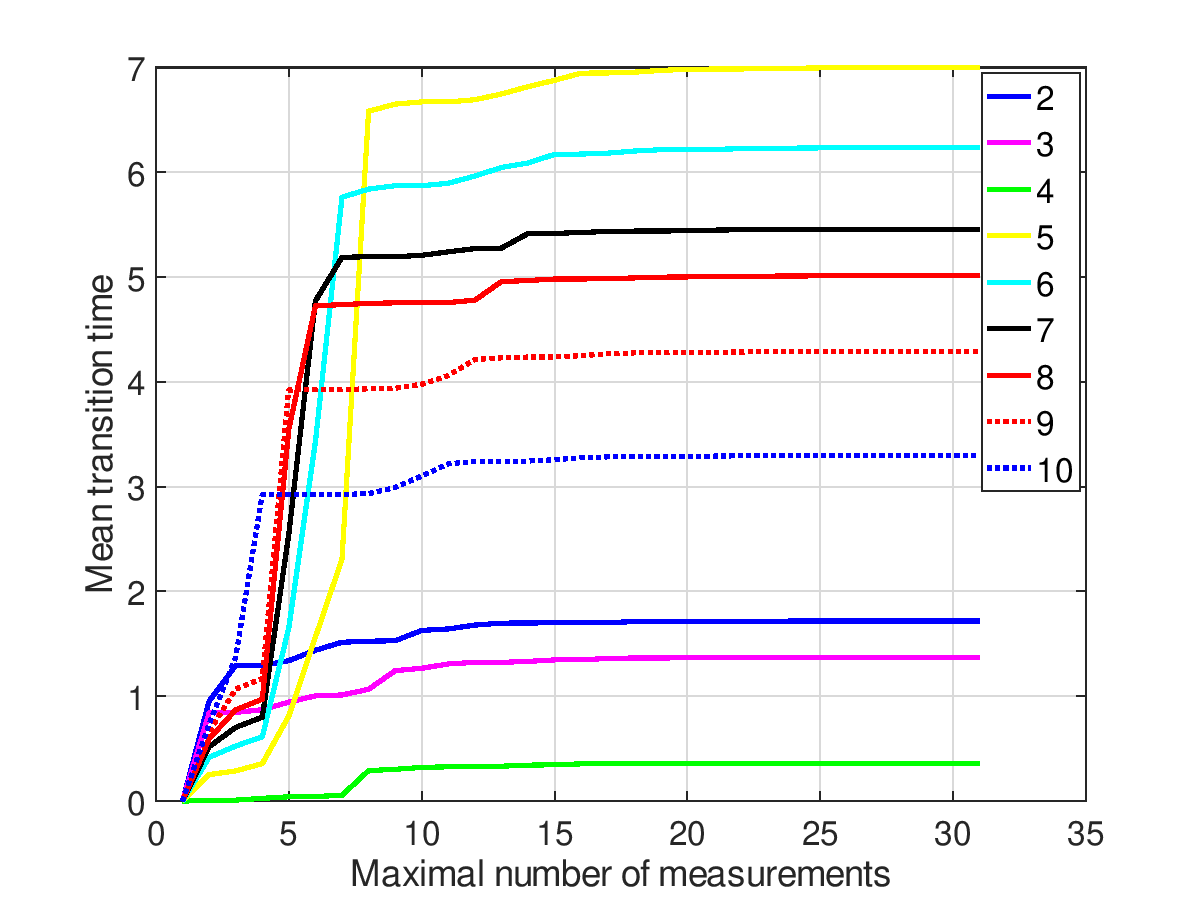}
    \caption{Mean transition time $|\br_j\rangle\to|\br_5\rangle$ with $J\tau=\pi\hbar/2$
    a) in the localized basis, where $j=1,2,3,4$ and $j=7,9$ are equal
    and b) in the plane-wave basis. 
           }    
\label{fig:mfdt3}
\end{figure}

\section{Discussion and examples}
\label{sect:discussion}

In Sects. \ref{sect:properties} and \ref{sect:loc_states} we have found 
that degenerate eigenvalues of the unitary evolution operator $\exp(-iH\tau)$ 
and EOSs create eigenvalues of the monitored evolution matrix ${\hat T}_M$ 
on the unit circle. In particular, 
if the measured state $|\br_M\rangle$ is orthogonal to $n$ energy eigenstates, the
corresponding eigenvalues $\lambda_j$ of the monitored evolution matrix ${\hat T}_M$ 
obey $|\lambda_j|=1$.
These results are in agreement with previous calculations based on the classical 
electrostatic
picture and indicate that the poles of the electrostatic potential correspond with the eigenvalues
of ${\hat T}_M$. For instance, it is known that near a spectral degeneracy or in the presence of an EOS one 
or more poles of the electrostatic potenial move towards the unit 
circle~\cite{PhysRevResearch.1.033086,PhysRevResearch.2.043107,PhysRevResearch.4.023129}. 
The connection of the poles and the eigenvalues originates in the sum
$G(z)=\sum_{m\ge 1}[{\hat T}_M/z]^{m-1}=z(z{\bf 1}-{\hat T}_M)^{-1}$, since the poles
of $G(z)$ are the eigenvalues of ${\hat T}_M$.

The results for the transition amplitude can be used to 
calculate the transition probability of the unitary evolution ${\bar P}_{kk'}$ of Eq. (\ref{time_av}) 
as well as the monitored transition probability $|\phi_{MM'}(m,\tau)|^2$ obtained from Eq. (\ref{qws_loc2}).
Then the mean transition time ${\bar\tau}_{{\cal N};MM'}$ of Eq. (\ref{mfdt00}) 
can be calculated as for specific examples. 
Since $\phi_{MM\rq{}}(m,\tau)=0$ for $m>m_f$, this expression saturates when ${\cal N}$ approaches $m_f$. 
We will focus here on a specific localized basis and plane waves as a delocalized basis to 
reveal some properties which distinguish these two cases. In order to obtain a robust result for
the mean transition time we need a sufficiently large number of measurements ${\cal N}$,
whose value depends essentially on the eigenvalues of $T_M$. The smaller their absolute value is, the
faster the transition probability $|\phi_{MM\rq{}}(m,\tau)|^2$ decays with $m$. 
Near spectral degeneracies $m_f$ can become arbitrarily large, 
which means that ${\cal N}$ needs to be very large for reliable results of the mean transition time. 
To demonstrate these effects we consider the linear spectrum $E_k=J(k-1)$,
which yields a spectral degeneracy for the evolution when $(E_k-E_l)\tau=J\tau(k-l)$ 
is a multiple of $2\pi\hbar$.

Next, we define a specific basis, namely a basis with
a single delocalized vector $\bq_1$, which always exists due to detailed balance according to
Eq. (\ref{detailed_balance}), 
and orthogonal localized vectors $\bq_k=(q_{1,k},\ldots,q_{N,k})^T$ with
(cf. App. \ref{app:local_basis})
\beq
\label{loc_basis2}
q_{M,1}=1/\sqrt{N}
,\ \
q_{M,k}=\frac{1}{\sqrt{k(k-1)}}\cases{
0 & $1<k< M\le N$ \cr
-k+1 & $1<k=M\le N$ \cr
1 & $1\le M<k\le N$ \cr
}
,
\eeq
where $\{\bq_1,\ldots,\bq_N\}$ forms an orthonormal basis. 
This basis is inspired by an eigenbasis of the Hamiltonian that connects
all $N$ vertices of a graph with equal weights~\cite{Ziegler_2024}.

The delocalized vector $\bv_1$ gives $c_1=1/N$, while the other vectors $\bq_{k}$ ($2\le k\le N$)
are localized with the inverse participation ratio
\beq
\label{loc_criterion1}
c_k
=\frac{1}{N^2}+(1-1/k)^2+\sum_{j=k+1}^{N}\frac{1}{j^2(j-1)^2}
\eeq
with the right-hand side bounded from below by $1/4$ for any $N$.
Within this basis an energy eigenstate is expressed in the basis $\{|\br_k\rangle\}$ as
$|E_j\rangle=\sum_{k}q_{k,j}|\br_k\rangle$ with
\beq
\label{energy_state}
|E_1\rangle=\frac{1}{\sqrt{N}}\sum_{k=1}^N|\br_k\rangle
\ , \
|E_j\rangle=-\sqrt{1-1/j}|\br_j\rangle+\frac{1}{\sqrt{j(j-1)}}\sum_{k=1}^{j-1}|\br_k\rangle
\ \ (2\le j\le N),
\eeq
where $|E_1\rangle$ is delocalized on the graph and $|E_j\rangle$ ($2\le j\le N$) is
localized at $|\br_j\rangle$.

\subsection{Strong localization}
\label{sect:strong_loc}

For the basis defined in Eq. (\ref{loc_basis2}) the strongest localization effect
on the QWS exists for $M=N$, where the summation 
in Eq. (\ref{qws_loc}) is reduced to $j_m=1,N$ with $q_{N,1}=1/\sqrt{N}$ 
and $q_{N,N}=\sqrt{1-1/N}$. This represents an effective two-level system
with transitions between $|\br_{1}\rangle$ and $|\br_N\rangle$, 
whose projected evolution matrix reads
\beq
\Pi_{{\cal L}_N}{\hat T}_N\Pi_{{\cal L}_N}
=\frac{1}{N}\pmatrix{
N-1 & z_N\sqrt{N-1} \cr
z_N\sqrt{N-1} & z_N^2 \cr
}
\eeq
with eigenvalues $\lambda_1=0$ and $\lambda_2=(z_N^2+N-1)/N$.
The decay of the transition amplitude is determined by
$|\lambda_2|^2=(1-1/N)^2+1/N^2+2(1-1/N)\cos(E_N\tau)/N$, which
decreases with increasing $N$. The asymptotic behavior 
scales with $N$ as
\beq
|\lambda_2|^{m-1}=\exp\left\{-\frac{m-1}{N}\left[1-\cos(E_N\tau)\right]+O(N^{-2})\right\}
.
\eeq
The transition amplitude $\phi_{MM'}(m,\tau)$ also depends on $q_{M',j}$, where
$q_{M',1}=1/\sqrt{N}$ and 
\beq
\label{relation1}
q_{M',N}=\cases{
1/\sqrt{N(N-1)} & for $M'<N$ \cr
-\sqrt{1-1/N} & for $M'=N$ \cr
}
\eeq
contribute to $M=N$. Thus, for large $N$ the amplitude for the transition 
$|\br_N\rangle\to|\br_N\rangle$ is of order 1, whereas the amplitudes
of other transitions
are at most of order $1/N$.

\subsection{Transition probability for the unitary evolution}

First, we consider the unitary evolution for the basis defined in Eq. (\ref{loc_basis2})
and calculate the right-hand side of Eq. (\ref{time_av}):
\[
{\bar P}_{1k\rq{}}={\bar P}_{k1}=\frac{1}{N}
,
\]
which means that the transition probability from or to the state $|\br_1\rangle$ is equal $1/N$.
This case is special because $k=1$ or $k\rq{}=1$ refer to the uniformly delocalized 
eigenvector $\bv_1$ due to the detailed balance condition in Eq. (\ref{detailed_balance}).
In contrast, for the localized states with $2\le k,l\le N$ the transition probability 
${\bar P}_{kl}$ decays with increasing $k$ and $l$.
Since ${\bar P}_{kl}$ is symmetric, we can assume without restriction that $l\ge k$.
Then we get
\beq
\label{ue_transition}
{\bar P}_{kl}
=\sum_{j=1}^N|q_{k,j}|^2|q_{l,j}|^2
=\frac{1}{N^2}+\frac{1}{l^2}
+\sum_{j=l+1}^{N}\frac{1}{j^2(j-1)^2} 
\ \ (2\le k<l<N),
\ \ \ 
 {\bar P}_{kN}=\frac{2}{N^2}\ \ (2\le k<N)
.
\eeq
In contrast, for the plane-wave basis of Eq. (\ref{plane_wave_basis}) 
there is no decay with $l$ but 
${\bar P}_{kl}=1/N$. Thus, the transition between any pair of states
on the graph has the same probability $1/N$ after time average.

\subsection{Examples for a graph of dimension $N=10$}

In Figs. \ref{fig:decay} -- \ref{fig:mfdt3} we visualize some results
of a quantum walk on a graph with $N=10$ sites and
with matrix elements $H_{kl}=J\sum_{j=1}^{10} q_{k,j}(j-1)q_{l,j}^*$
of the Hamiltonian $H$ in the basis $\{|\br_k\rangle\}$. 
The localization effect on the spatial structure of the transition 
probability is illustrated for the unitary evolution in Fig. \ref{fig:decay}a
and for the monitored evolution in Fig. \ref{fig:decay}b-c. 
Fig. \ref{fig:decay}a visualizes the typical localization effect
with a monotonic decay of ${\bar P}_{kl}$ with the distance $|k-l|$.
On the other hand, the behavior of the monitored transition probability
$\sum_{m=1}^{500}|\phi_{MM'}(m,\tau)|^2$ is more complex and depends strongly
on the parameter $J\tau$.
For instance, it is similar to the unitary evolution for small $J\tau$ due to
the quantum Zeno effect (Fig. \ref{fig:decay}b). This means that the system remains
near the initial state due to frequent measurements.
For $J\tau=\hbar$ the transition probability
is flat for $M'<M$ at a value less than $1$, typically of order $1/N$, 
and jumps to 1 for $M'\ge M$ (Fig. \ref{fig:decay}b). For $J\tau=\pi\hbar/2$ it is smaller
than 1 away from the line $M'=M$ in Fig. \ref{fig:decay}d. All three cases have in common
that the monitored transition probability is 1 if $M'=M$, which reflects that the return probability
is always 1. The triangular structure is caused by the vanishing projected evolution 
matrix elements $[\Pi_{{\cal L}_M}{\hat T}_M\Pi_{{\cal L}_M}]_{kl}=0$ for $2\le k,l<M$
and $q_{M',l}=0$ for $2\le l<M'$.

The quantum walks on the localized basis provide a 
structure of the graph with connections between the vertex states $\{|\br_k\rangle\}$
through the transition probabilities. While the unitary evolution
creates symmetric transitions in Fig. \ref{fig:decay}a, 
the monitored evolution creates asymmetric transitions in Fig. \ref{fig:decay}b-c
such that the quantum walk on the graph is directed.
The asymmetry is very weak in the quantum Zeno regime of Fig. \ref{fig:decay}b,
regardless if a localized or a plane-wave basis is used. Here only the result of the
localized basis is plotted but it is very similar for the plane-wave basis.
With less frequent measurements in Figs. \ref{fig:decay}c,d the asymmetry becomes
stronger. In general, the complex behavior is the result of the competition between localization and
measurements.
This is also reflected by the irregular behavior of the transition probability 
$|\phi_{MM\rq{}}(m,\tau)|^2$ as a function of the number of measurements $m$ in 
Fig. \ref{fig:tprob}a.

Finally, we analyze the mean transition time of the monitored evolution of Eq. (\ref{mfdt00})
for parameter values (i) $J\tau=\hbar$, which has no degeneracy, 
while for (ii) $(\pi/2+0.002)\hbar$ we are very close to multiple degeneracies of $E_k$. 
This is compared with the behavior for (iii) exactly at the degeneracies $J\tau=\pi\hbar/2$.
The eigenvalues $\{\nu_j\}$ of the monitored
transition matrix ${\hat T}_M$ are plotted in Fig. \ref{fig:mfdt1}a for case (i) with the localized as 
well as with the plane-wave basis. The corresponding eigenvalues for case (ii) are plotted in
Fig. \ref{fig:mfdt2}a. In both cases the typical eigenvalues are closer to the unit circle for
the localized states. Moreover, the tendency to degenerate is stronger for the eigenvalues of case (ii).
A few examples for the transition probabilities in Fig. \ref{fig:tprob} illustrate that
their distributions with respect to the number of measurements is broader and smoother for
the localized basis.

The mean transition time for the transitions $|\br_{M\rq{}}\rangle\to|\br_M\rangle$
as a function of the maximal number of measurements ${\cal N}$,
defined in Eq. (\ref{mfdt00}), is plotted in Fig. \ref{fig:mfdt1}c,d for the case (i) 
and in Fig. \ref{fig:mfdt2}c,d for the case (ii). For the localized basis 
the shortest mean transition time appears
for the return transition $|\br_5\rangle\to|\br_5\rangle$ in Fig. \ref{fig:mfdt1}c
as well as in Figs. \ref{fig:mfdt2}c, \ref{fig:mfdt3}a, which is always an integer
due to the protected winding number of the return~\cite{Gruenbaum2013,quancheng20,Ziegler_2021}.
Moreover, the curves of
the mean transition times are equal for $|\br_{1,2,3,4}\rangle\to|\br_5\rangle$ in
Figs. \ref{fig:mfdt1}c, \ref{fig:mfdt2}c, \ref{fig:mfdt3}a, as a consequence of
the relation in Eq. (\ref{relation2}), and they are equal in Fig. \ref{fig:mfdt3}a
for $|\br_{7,9}\rangle\to|\br_5\rangle$ as a consequence of $E_7\tau=3\pi\hbar$ and $E_9\tau=4\pi\hbar$.
This is different for the plane-wave basis, where the shortest
mean transition time appears for the transition $|\br_4\rangle\to|\br_5\rangle$ in 
Figs. \ref{fig:mfdt1}d, \ref{fig:mfdt2}d and \ref{fig:mfdt3}b.
This behavior reflects the localization effect on the mean transition time. The vicinity of degeneracies
in case (ii) leads to an increasing mean transition time for several transitions, as visualized
in Fig. \ref{fig:mfdt2}c,d. Remarkably, the diverging mean transition times do not appear exactly
at the degeneracy point, as demonstrated in Fig. \ref{fig:mfdt3}.
The mean transition time is typically shorter for the plane-wave basis, indicating that the localized
states prevents the quantum walk to perform the transition directly.

The behavior of the mean transition time with respect to the maximal number of measurements ${\cal N}$
is much smoother for the localized basis than for the plane-wave basis, where the latter jumps in
steps (cf. Figs. \ref{fig:mfdt1}b and \ref{fig:mfdt3}b). This indicates that the transition
probability $|\phi_{MM\rq{}}(m,\tau)|^2$ does not decay smoothly with $m$ but has large
contributions for certain values of $m$ (cf. Fig. \ref{fig:tprob}b).
On the other hand, Fig. \ref{fig:tprob}a illustrates that it requires more measurements to 
reach $m_f$ in the localized basis.


\section{Conclusions}

Our approach to the monitored evolution on a graph with basis $\{|\br_k\rangle\}$
can be summarized as the mapping from energy levels $\{E_j\}$ and spectral weights
$\{q_{k,j}\equiv\langle\br_k|E_j\rangle\}$ to the monitored evolution matrix
${\hat T}_M$ as $\{E_j,q_{k,j}\}_{k,j=1,\ldots N}\to {\hat T}_M$. The corresponding
unitary evolution is defined by the unitary evolution matrix
$\sum_j e^{-iE_j t}q_{k,j}q^*_{l,j}$ for the transition $|\br_l\rangle\to|\br_k\rangle$
during the time $t$. This enabled us to calculate the transition probabilities and
the mean transition times.

Localization is defined through the inverse participation ratio,
which is identified in the unitary evolution by the time-averaged
return probability ${\bar P}_{kk}$.
In particular, localized states are created by $q_{M,j}=0$ for $j\in {\cal L}_M$. 
Then only the evolution matrix projected onto $j\notin {\cal L}_M$
provides the effective monitored evolution. The monitored evolution represents directed 
quantum walks on the graph through asymmetric transition probabilities. In contrast,
the unitary evolution has always symmetric transition probabilities.
Localized states as well as a degeneracy $(E_k-E_l)\tau=0\ ({\rm mod}\ 2\pi)$ create eigenstates
of the monitored evolution matrix ${\hat T}_M$ whose eigenvalues are on the unit circle.

The mean transition time enables us to distinguish between a localized and delocalized
basis. In our examples we have observed that a localized basis, in comparison to 
a delocalized basis, has substantially larger mean transition times, implying that
$m_f$ is typically smaller for the plane-wave basis. Moreover, a localized basis 
favors a quicker return to the initial state in comparison with any transition
to other states.
Moreover, the reduction of the Hilbert space by localization may lead to identical 
mean transition times for different transitions.

The competition of localization and projective measurements makes the monitored 
transition probability quite sensitive to changes of the spectral parameter $J\tau$.
This is also reflected by the location of the eigenvalues of ${\hat T}_M$ on the
complex unit disk. Since the distribution of these eigenvalues depend on the chosen basis,
it would be interesting to determine their correlation in a quantative way,
similar to the eigenvalue correlation in the random matrix 
theory~\cite{1966NucPh..78R.696L,10.2307/2027409,mehta2004random}.
\vs

\no
{\bf Acknowledgment}

I am grateful to Eli Barkei, Quancheng Liu and Sabine Tornow for enlightening discussions.

\appendix

\section{Monitored evolution}
\label{app:qws}

Starting from the expression in Eq. (\ref{trans_amp00}) we can write with
$\Pi_M={\bf 1}-|\br_M\rangle\langle\br_M|$ for the monitored transition amplitude 
\beq
\label{trans_amp01a}
\phi_{MM\rq{}}(m,\tau)=\langle\br_M|e^{-iH\tau}\left(\Pi_M e^{-iH\tau}\right)^{m-1}|
\br_{M\rq{}}\rangle
\]
\[
=\sum_{k_1,k_2,\ldots,k_{m-1}\ne M}\langle\br_M|e^{-iH\tau}|\br_{k_1}\rangle
\langle\br_{k_1}|e^{-iH\tau}|\br_{k_2}\rangle\cdots\langle\br_{k_{m-1}}|e^{-iH\tau}|\br_{M\rq{}}\rangle
\]
\[
=\sum_{k_1,k_2,\ldots,k_{m-1}\ne M}\sum_{j_1,j_2,\ldots,j_m}
e^{-i(E_{j_1}+E_{j_2}+\cdots +E_{j_m})\tau}
q_{M,j_1}q^*_{k_1,j_1}q_{k_1,j_2}q^*_{k_2,j_2}\cdots q_{k_{m-1},j_m}q^*_{M\rq{},j_m}
.
\eeq
After reordering the summations we get
\beq
\phi_{MM\rq{}}(m,\tau)
=\sum_{j_1,j_2,\ldots,j_m}e^{-i(E_{j_1}+E_{j_2}+\cdots +E_{j_m})\tau}
q_{M,j_1}K_{j_1,j_2}\cdots K_{j_{m-1},j_m}q^*_{M\rq{},j_m}
\]
with
\[
K_{j_l,j_{l+1}}:=\sum_{k_l\ne M}q^*_{k_l,j_l}q_{k_l,j_{l+1}}
=\delta_{j_l,j_{l+1}}-q^*_{M,j_l}q_{M,j_{l+1}}
,
\eeq
which is the kernel defined already in Eq. (\ref{energy_basis}). 
This expression is reminiscent of a functional integral~\cite{glimm2012quantum,negele2019quantum}.
But instead of an integration we have a summation here with discrete time steps $\tau$. 
Therefore, we will call it a quantum walk sum, which is much easier to calculate than Feynman\rq{}s 
path integral, especially when we employ numerical methods.
With this representation of the monitored transition amplitude the effect of the energy orthogonal states 
$|\br_k\rangle$ with $q_{k,j}=0$ is directly visible, since these states reduce the summation with
respect to energy states $\{E_j\}$ with $q_{k,j}\ne 0$.  

\section{Degeneracy and eigenvectors of the evolution}
\label{app:degeneracy}

The eigenvalue equation (\ref{ONS}) implies the relation
\beq
\label{eigenvalue2}
\sum_l
{\hat T}_{M;kl}q^*_{M',l}=(1-\delta_{MM'})z_k^2q^*_{M',k}
\ \ {\rm with}\ \
q^*_{M',l} =z_l^*q_{M',l}^*
,
\eeq
where $\{\bQ_M\}$ with $\bQ_M=(q_{M,1},\ldots, q_{M,N})^T$
is an orthonormal basis: $\sum_kq^{}_{M,k}q^*_{M',k}=\delta_{MM'}$.
Thus, the mapping of the vector $\bQ^*_{M\rq{}}$ ($M\rq{}\ne M$) by $T_M$ is unitary and, 
therefore, does not change the length of ${\bf Q}^*_{M\rq{}}$.
In other words, we have ${\hat T}_M{\bf Q}^*_{M\rq{}}=D{\bf Q}^*_{M\rq{}}$ for all $M\rq{}\ne M$,
which is degenerated with respect to $M\rq{}$, such that any linear combination of
$\{{\bf Q}^*_{M'}\}$ satisfies the same equation. Introducing the Cartesian vector
$\be_k=(\langle E_1|E_k\rangle,\ldots,\langle E_N|E_k\rangle)^T$ and assuming that $z_j=z_k$,
we can define a linear combination for $\{{\bf Q}^*_{M'}\}_{M\rq{}\ne M}$ that is (i)
proportional to $\alpha \be_j+\beta \be_k$ and (ii) it is orthogonal to ${\bf Q}^*_M$.
It is sufficient to determine the coefficients $\alpha$, $\beta$ such that they satisfy condition (ii).
With the Cartesian vector $\be_j$ and the vector ${\bf Q}^*_{k}$ of Eq. (\ref{eigenvalue2})
we get $\be_j\cdot{\bf Q}^*_{k}=z_j^*q^*_{k,j}$ and
\[
(\alpha\be_j+\beta\be_k)\cdot{\bf Q}^*_M=\alpha z_j^*q^*_{M,j}+\beta z_k^*q^*_{M,k}
,
\]
which vanishes for $\alpha=-\beta z_k^*q^*_{M,k}/z_j^*q^*_{M,j}$, provided that 
$z_j^*q^*_{M,j}\ne 0$.
This means that $\alpha\be_j+\beta\be_k$ lives in the space orthogonal to ${\bf Q}^*_M$.
Thus, the degeneracy  $z_j=z_k$ implies the eigenvalue equation for ${\hat T}_M$
\beq
\label{eigenvalue3}
{\hat T}_M(\alpha \be_j+\beta \be_k)=z_k^2(\alpha \be_j+\beta \be_k)
\eeq
with eigenvalue $z_k^2$, where $\alpha \be_j+\beta \be_k$ lives by definition in the space orthogonal 
to ${\bf Q}^*_M$.
This means that this eigenvector does not decay with the number of measurements in the 
monitored evolution.
The same argument can be applied to the case with more than two degenerate $z_j$.
Thus, degeneracies in terms of $E_k\tau\ ({\rm mod}\ 2\pi)$ have a strong effect on the
quantum walk by creating eigenstates of ${\hat T}_M$ with eigenvalues on the unit 
circle. 

\section{Recursion relation}
\label{app:recursion}

The monitored transition amplitude of Eq. (\ref{trans_amp01a}) reads
\[
\langle \br_M|e^{-iH\tau}[({\bf 1}-|\br_M\rangle\langle\br_M|)e^{-iH\tau}]^{m-1}|\br_{M\rq{}}\rangle
\]
\[
=\langle \br_M|e^{-2iH\tau}[({\bf 1}-|\br_M\rangle\langle\br_M|)e^{-iH\tau}]^{m-2}|\br_{M\rq{}}\rangle
-\langle \br_M|e^{-iH\tau}|\br_M\rangle
\langle \br_M|e^{-iH\tau}[({\bf 1}-|\br_M\rangle\langle\br_M|)e^{-iH\tau}]^{m-2}|\br_{M\rq{}}\rangle
\]
\[
=\sum_{k\ne M}
\langle \br_M|e^{-iH\tau}|\br_k\rangle
\langle \br_k|e^{-iH\tau}[({\bf 1}-|\br_M\rangle\langle\br_M|)e^{-iH\tau}]^{m-2}|\br_{M\rq{}}\rangle
,
\]
which also reads
\beq
\label{recursion00}
\phi_{MM\rq{}}(m,\tau)=\sum_{k\ne M}
\langle \br_M|e^{-iH\tau}|\br_k\rangle\phi_{kM\rq{}}(m-1,\tau)
\eeq
with $\phi_{kM\rq{}}(m-1,\tau)
=\langle \br_k|e^{-iH\tau}[({\bf 1}-|\br_M\rangle\langle\br_M|)e^{-iH\tau}]^{m-2}|\br_{M\rq{}}\rangle$.
For $m=2$ this yields
\[
\phi_{MM\rq{}}(2,\tau)
=\sum_{k\ne M}\langle \br_M|e^{-iH\tau}|\br_k\rangle\langle \br_k|e^{-iH\tau}|\br_{M\rq{}}\rangle
.
\]

\subsection{Stationary solution}
\label{app:stationary}

A stationary solution of Eq. (\ref{recursion2}) would obey the equation
\beq
\label{stat_cond}
\sum_{l\rq{}}\left[
\delta_{ll\rq{}}-\langle \br_l|e^{-iH\tau}|\br_{l\rq{}}\rangle(1-\delta_{l\rq{}M})\right]
{\bar\phi}_{l\rq{}k}=0
.
\eeq
This equation is solved either by ${\bar\phi}_{l\rq{}k}=0$, which corresponds with a decaying
transition amplitude, or if $({\bar\phi}_{1k},\ldots,{\bar\phi}_{Nk})^T$ is an eigenvector 
with zero eigenvalue. When we expand this vector in the Cartesian basis as $\sum_l{\bar\phi}_{lk}\be_l$,
we get
\beq
\label{stat_sol2}
{\bar\phi}_{Mk}\be_M+({\bf 1}-U)\sum_{l\ne M}{\bar\phi}_{lk}\be_l=0
\ ,\ \ \
U_{ll\rq{}}=\langle \br_l|e^{-iH\tau}|\br_{l\rq{}}\rangle=\sum_jz^2_jq_{l,j}q^*_{l\rq{},j}
.
\eeq
A solution is ${\bar\phi}_{Mk}=0$ and ${\bf 1}-U$ must have a vanishing eigenvalue. Moreover, the
corresponding eigenvector must be of the form $\sum_{k\ne M}\alpha_k\be_k$ with appropriate
coefficients $\{\alpha_k$\}. After replacing 
$({\bar\phi}_{1k},\ldots,{\bar\phi}_{Nk})^T\to (q_{1,k},\ldots,q_{N,k})^T$ 
we get from Eq. (\ref{stat_sol2}) the relation
\beq
\sum_{j}\left(\delta_{lj}-\sum_{l\rq{}}q_{l,l\rq{}}z^2_{l\rq{}}q^*_{j,l\rq{}}\right)q_{j,k}
=(1-z_k^2)q_{l,k}
,
\eeq
which vanishes for a $z_k^2=1$. In the latter case we have an eigenvector that satisfies the stationary
condition in Eq. (\ref{stat_cond}) when this eigenvector is orthogonal to $\be_M$ 
(i.e., when $q_{M,k}=0$). Therefore, we have a stationary solution of the recursion 
relation (\ref{recursion2}) when there is an $k$ with $z_k^2=1$ and $q_{M,k}=0$.

\section{A special localized basis}
\label{app:local_basis}

A basis with
a single delocalized vector $\bv_1$, which always exists due to detailed balance according to 
Eq. (\ref{detailed_balance}), and orthogonal localized vectors
$\bv_M=(q_{1,M},\ldots,q_{N,M})^T$ with
\beq
\label{loc_basis2a}
q_{k,l}
=\frac{1}{\sqrt{N}}\delta_{l1}
+(1-\delta_{l1})\frac{1}{\sqrt{l(l-1)}}\left[(-l+1)\delta_{kl}+\Theta_{kl}\right]
\ {\rm with}\ \ 
\Theta_{kl}=\cases{
1 & $k< l$ \cr
0 & $k\ge l$ \cr
} 
.
\eeq
This gives for $1\le M\le N$
\beq
\label{loc_basis2c}
q_{M,1}=1/\sqrt{N}
,\ \
q_{M,k}=\frac{1}{\sqrt{k(k-1)}}\cases{
0 & $1<k<M\le N$ \cr
-k+1 & $1< k=M\le N$ \cr
1 & $1\le M< k\le N$ \cr
}
\eeq
such that from Eq. (\ref{loc_basis2}) we get for the kernel
\[
K_{M;kl}=\delta_{kl}-q_{M,k}^*q^{}_{M,l}
=\cases{
\delta_{kl}-1/N & for $k,l=1$ \cr
A_{kl} & for $1<k,l\le M$ \cr
\delta_{kl} & for $M< k\le N$ or $M< l\le N$\cr
}
,
\]
where $A_{kl}$ is nonzero for $k,l\le M$. This implies for $2\le M<k,l\le N$
\beq
\langle E_k|T_M^{m-1}|E_l\rangle=\langle E_l|T_M^{m-1}|E_k\rangle
=\delta_{kl}e^{-iE_k\tau (m-1)}
.
\eeq
Thus, the state $|E_M\rangle$ is dark for any of the initial states $\{|E_l\rangle\}_{M<l}$.
This represents a special case of Eq. (\ref{local1}) for the basis in Eq. (\ref{loc_basis2}).



\end{document}